\documentclass[%
 aip,
 amsmath,amssymb,
 reprint,%
]{revtex4-1}

\usepackage{graphicx}% Include figure files
\usepackage{dcolumn}% Align table columns on decimal point
\usepackage{bm}% bold math
\usepackage{hyperref}% add hypertext capabilities
\DeclareMathAlphabet{\altmathcal}{OMS}{cmsy}{m}{n}

\usepackage[utf8]{inputenc}
\usepackage[T1]{fontenc}
\usepackage{mathptmx}
\usepackage{etoolbox}
\usepackage{amsmath}
\usepackage{amssymb}

\makeatletter
\def\@email#1#2{%
 \endgroup
 \patchcmd{\titleblock@produce}
  {\frontmatter@RRAPformat}
  {\frontmatter@RRAPformat{\produce@RRAP{*#1\href{mailto:#2}{#2}}}\frontmatter@RRAPformat}
  {}{}
}%
\makeatother

\hypersetup{
  colorlinks   = true, %Colours links instead of ugly boxes
  urlcolor     = blue, %Colour for external hyperlinks
  linkcolor    = blue, %Colour of internal links
  citecolor   = blue %Colour of citations
}

\begin{document}

%\preprint{AIP/123-QED}

\title{Mean-field model of synchronization for open-loop, swirl controlled thermoacoustic system}
% Force line breaks with \\
\author{Samarjeet Singh} \altaffiliation[]{These authors contributed equally to this work.} \email{samarjeet.singh448@gmail.com.}
\affiliation{Department of Aerospace Engineering, Indian Institute of Technology Madras, Tamil Nadu 600036, India}  
 
\author{Ankit Kumar Dutta}%
\altaffiliation[]{These authors contributed equally to this work.}
\affiliation{Department of Aerospace Engineering, Indian Institute of Science, Bangalore 560012, India}

\author{Jayesh M. Dhadphale}
\affiliation{Department of Aerospace Engineering, Indian Institute of Technology Madras, Tamil Nadu 600036, India} 

\author{Amitesh Roy}
\altaffiliation[]{Presently at Institute for Aerospace Studies, University of Toronto, Ontario M3H 5T6, Canada.}
\affiliation{Department of Aerospace Engineering, Indian Institute of Technology Madras, Tamil Nadu 600036, India}
%\affiliation{Institute for Aerospace Studies, University of Toronto, Ontario M3H 5T6, Canada} 

\author{R. I. Sujith}
\affiliation{Department of Aerospace Engineering, Indian Institute of Technology Madras, Tamil Nadu 600036, India} 

\author{Swetaprovo Chaudhuri}
\affiliation{Institute for Aerospace Studies, University of Toronto, Ontario M3H 5T6, Canada}

\date{\today}% It is always \today, today,

\begin{abstract}
%\textbf{Punchline:} We develop a theoretical model for open-loop control experiments suppressing thermoacoustic instability. The model only need the heat release rate spectrum at "stable" operation as a input to reproduce the features of temporal and spatio-temporal dynamics during all the states observed in the experiments.

Open-loop control is known to be an effective strategy for controlling self-excited periodic oscillations in turbulent combustors. The occurrence of thermoacoustic instability is caused by a positive feedback between the unsteady heat release rate and the acoustic modes of the combustor leading to pressure oscillations having ruinously large amplitude. Here, we present experimental observations and a synchronization model for the suppression of thermoacoustic instability achieved by rotating the otherwise static swirler in a lab-scale turbulent combustor. Using the model, we demonstrate and analyze the intermittent dynamics that emerge en route to the suppression of thermoacoustic instability. Starting with the state of thermoacoustic instability in the combustor, we find that a progressive increase in the swirler rotation rate leads to a transition from the state of limit cycle oscillations to the low amplitude aperiodic oscillations through a state of intermittency. To model such a transition while also quantifying the underlying synchronization characteristics, we extend the model of Dutta et al. [\href{https://doi.org/10.1103/PhysRevE.99.032215}{Phys. Rev. E \textbf{99}, 032215 (2019)}] by introducing a feedback between the ensemble of phase oscillators and the acoustic. The coupling strength in the model is determined by considering the effect of the acoustic and swirl frequencies. The assumption that coupling strength among the oscillators is a linear combination of acoustic and swirler rotation frequency is justified \textit{a posteriori}. The link between the model and experimental results is quantitatively established by implementing an optimization algorithm for model parameter estimation. We show that the model is capable of replicating the bifurcation characteristics, features of time series, probability density function, and amplitude spectrum of acoustic pressure and heat release rate fluctuations at various dynamical states observed during the transition to the state of suppression. Most importantly, we discuss the flame dynamics and demonstrate that the model without any spatial inputs qualitatively captures the characteristics of the spatio-temporal synchronization between the local heat release rate fluctuations and the acoustic pressure that underpins a transition to the state of suppression. As a result, the model emerges as a powerful tool for explaining and controlling instabilities in thermoacoustic and other extended fluid dynamical systems, such as aeroelastic and aeroacoustic systems, where spatio-temporal interactions lead to rich dynamical phenomena.

\end{abstract}

\maketitle

\begin{quotation}

The occurrence of undesirable high amplitude, self-sustained periodic acoustic oscillations observed in turbulent combustion systems is referred to as thermoacoustic instability. Thermoacoustic instability results from a positive feedback between the acoustic pressure and the heat release rate fluctuations. Mitigating such instability remains a challenge, even after decades of extensive research. Studies on suppressing thermoacoustic instability in turbulent combustion systems are important for developing mitigation strategies and control systems. In this paper, we demonstrate the suppression of self-excited thermoacoustic instability in a turbulent combustion system using an actuated swirler. We provide a model of synchronization to explain the mechanism of control. The model is built by considering the nonlinear response of the flame as an ensemble of phase oscillators, evolving collectively under the influence of acoustic pressure.  In order to suppress thermoacoustic instability, a positive feedback between the phase oscillators and the acoustic field is disrupted by increasing the swirler rotation rate. Therefore, we incorporate the effect of the acoustics and active swirler in the model, which plays a significant role in determining the flame response and hence, suppression of thermoacoustic instability. The model not only captures temporal dynamics but also reproduces the features of the spatio-temporal synchronization observed in the experiments. Finally, we show the relationship between the control parameter in the model and the experiments using a simple parameter estimation technique.

\end{quotation}

\section{\label{sec:intro}Introduction}

Lean combustion systems are prone to thermoacoustic instability wherein high amplitude acoustic pressure oscillations occur due to positive feedback between the unsteady turbulent flame and the acoustic modes of the combustion chamber \cite{lieuwen2005combustion,sujith2021thermoacoustic}. These oscillations can result in disastrous consequences such as severe damage to the engine components and even mission failures \citep{juniper2018sensitivity}. As a result, it is vital to develop methods to suppress thermoacoustic instability. The present paper discusses a novel strategy of controlling limit cycle oscillations by actuating the swirler used for flame stabilization. To explain the method of suppressing limit cycle oscillations, we provide a synchronization model.

Self-excited nonlinear thermoacoustic oscillations arise when the heat release rate oscillations are in phase with the acoustic pressure oscillations \citep{rayleigh1878explanation}, and the amount of energy added to the acoustic field due to the nonlinear feedback from the flame exceeds the acoustic damping in the combustor \citep{chu1965energy}. There are a number of physical mechanisms which generate heat release rate fluctuations; these include equivalence ratio fluctuations \citep{lieuwen1998role,kim2010response}, swirl number fluctuations \cite{komarek2010impact}, flame-vortex interactions \citep{poinsot1987vortex,renard2000dynamics}, and entropy fluctuations \cite{polifke2001constructive,goh2013influence}. Heat release rate fluctuations generated through these mechanisms undergo feedback coupling and mutual synchronization with acoustic pressure fluctuations in the confinement, resulting in the occurrence of thermoacoustic instability \cite{pawar2017thermoacoustic}. 

To comprehend the physical mechanism for the onset of thermoacoustic instability, it is important to investigate the spatio-temporal behavior of the thermoacoustic systems. The onset of periodic oscillations (thermoacoustic instability) occurs via a transition from the stable operation of the combustor (otherwise known as combustion noise) through intermittency. Intermittency is the dynamical state where bursts of periodic oscillations appear amidst epochs of aperiodic oscillations at apparently random intervals \citep{nair2014intermittency}. Recently, several studies have performed the spatio-temporal analysis of the synchronization between the local heat release rate fluctuations from the flame and the acoustic pressure oscillations \citep{mondal2017onset, pawar2019temporal, hashimoto2019spatiotemporal, roy2021flame}. They reported that when the control parameter is changed, the coupling between the acoustic field and the local heat release rate increases, which leads to the emergence of order from disorder during the transition to thermoacoustic instability. Consequently, mitigation strategies aim at interrupting the coupling between the acoustic pressure field and the unsteady heat release rate fluctuations. 

These control strategies are broadly classified as passive control \cite{richards2003passive, zhao2015review} and active control \cite{mcmanus1993review, zhao2018review}. In passive control, some aspects of the combustor, such as the acoustic characteristics or the heat release rate dynamics is changed independently of the combustor operation \citep{zhao2015review}. The change in acoustic characteristics can be achieved by the use of acoustic damping resonators \citep{dupere2005use} and the change in heat release dynamics by fuel injection strategy \citep{steele2000passive} or fuel staging \cite{samarasinghe2017effect}. However, passive control strategies are effective only over a limited range of frequencies, require expensive and time-consuming design adjustments, and may be detrimental to engine performance. An alternate technique is active control which can be implemented in two different ways \citep{candel2002combustion}: closed-loop and open-loop control. During closed-loop active control, the state of the combustor is continuously monitored and control measures are adopted based on the specific state of the system \citep{zhao2018review}. In contrast, open-loop active control is achieved by forcing the system using actuators, without any feedback from controllers or sensors monitoring the dynamics in the combustor \citep{cosic2012open}. Active control is typically achieved through external acoustic forcing \cite{bellows2008nonlinear, balusamy2015nonlinear} or fuel-air modulation \cite{seume1998application, uhm2005low} where the suppression depends on the forced synchronization of the thermoacoustic system \cite{balusamy2015nonlinear, guan2019open, roy2020mechanism}. However, the implementation of these techniques remains quite challenging. Specifically, acoustic forcing mechanisms do not scale up with the amplitude levels of realistic turbulent combustors, and modulating fuel-air values are unreliable over the extremely long lifespan of turbulent combustors \citep{wu2018numerical,guan2019open}. 

Given these challenges, Gopakumar et al. \cite{gopakumar2016mitigating} proposed a novel strategy involving the use of an actuated swirler for altering the coupling between the flame and the acoustic fluctuations. They found that actuating the static swirler, used for flame stabilization inside the combustion chamber, even at moderate rotation rates significantly altered the flow field and the flame structure \cite{mahesh2018instability}. This alternation in the flow field caused by actuating swirler was associated with the suppression of thermoacoustic instability, where low amplitude aperiodic oscillations are observed through an intermittency route, hinting towards the de-synchronization of the acoustic pressure and the heat release rate fluctuations during the state of suppression. The underlying mechanism of suppression was then modeled by Dutta et al. \cite{dutta2019investigating} based on the synchronization of flame oscillators through the Kuramoto model \citep{kuramoto1975self}, which was able to quantitatively capture the heat release rate response observed in the experiments. 

Despite the usefulness of the heat release rate model, the underlying mechanism by which the acoustic pressure is coupled was not included in the model. Therefore, in this study, we couple the acoustic elements with the heat release rate model to capture the dynamics observed during suppression experiments. We extend the model introduced by Dutta et al. \cite{dutta2019investigating} and build upon our recent work on the mean-field model of thermoacoustic transitions \citep{singh2022mean} to explain the method of suppression observed in the swirl-controlled experiments. The effect of the actuated swirler is included by accounting for the time scale of the swirler rotation. This mismatch between the convective time scale and the acoustic time scale is then shown to be responsible for weakening the feedback coupling between the heat release rate and acoustic pressure oscillations, leading to the de-synchronization between pressure and heat release rate fluctuations during the suppression of thermoacoustic instability. In this manner, we develop a thermoacoustic mean-field model for studying the transition to the state of suppression.

\begin{figure*}[t]
    \centering
    \includegraphics[width=\textwidth]{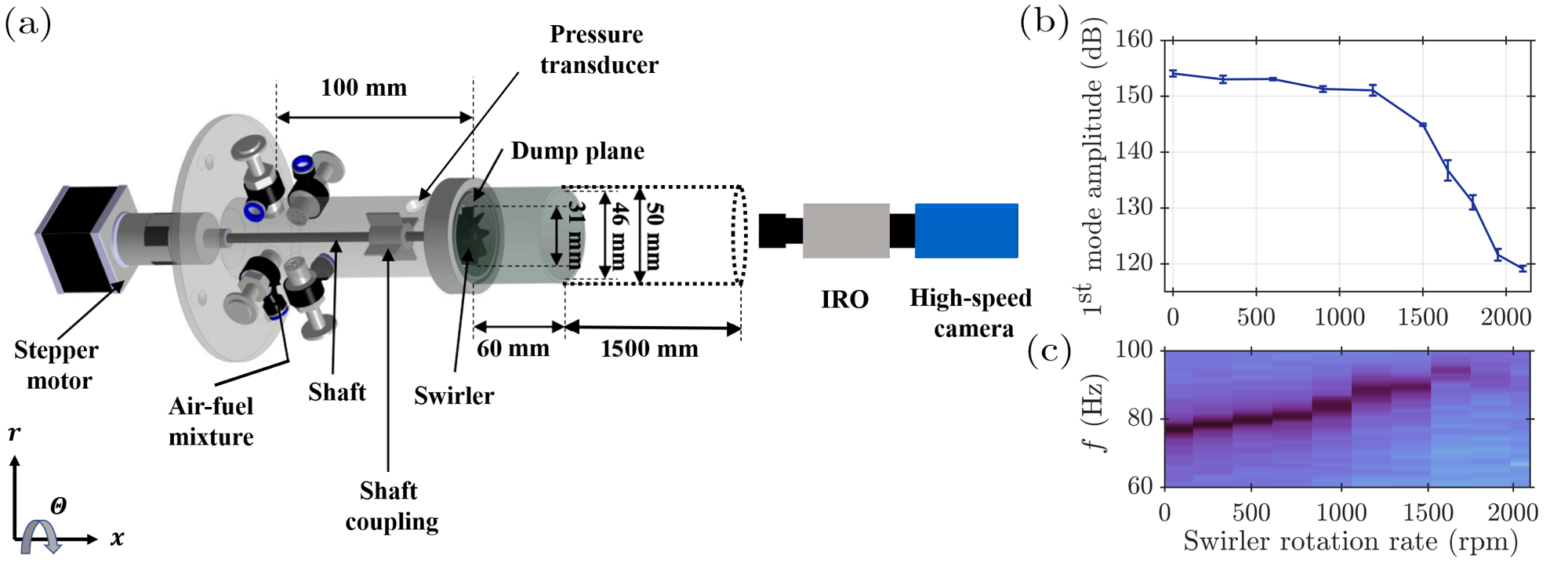}
    \caption{(a) A schematic of the experimental setup along with the diagnostic tools. (b) Variation of the acoustic pressure amplitude of the first dominant mode as a function of the swirler rotation rate, depicting the suppression of thermoacoustic instability. (c) The frequency corresponding to the acoustic mode as a function of the swirler rotation rate. The error bar in (b) is associated with the variation of the amplitudes across three experimental runs.}
    \label{fig:experiment}
\end{figure*}

Our model demonstrates the transition from thermoacoustic instability to the state of suppression by taking the amplitude spectrum of the heat release rate fluctuations as the only input from the experiments. After qualitatively confirming that a transition to the state of suppression produced from the model corresponds with the experiments, we use an optimization technique to determine the relationship between the control parameter in the model and the experiments. We then investigate the dynamics of the flame by computing the time average over pressure maxima and minima during the state of thermoacoustic instability, intermittency and suppression. Finally, we demonstrate that our model without any spatial inputs qualitatively captures features of spatio-temporal synchronization obtained from the experiments.

The rest of the paper is organized as follows. We introduce the experimental setup and measurements in Sec. ${\ref{sec:experi}}$. Further, in Sec. ${\ref{sec:model}}$, we derive a thermoacoustic mean-field model followed by estimating model parameters from the experiments. This is followed by results and discussion in Sec. ${\ref{sec:result}}$. Finally, in Sec. ${\ref{sec:conclu}}$ we summarize the major findings.

\section{\label{sec:experi}Experimental setup and measurements}

The schematic of the lab-scale premixed turbulent combustor is shown in Fig. \ref{fig:experiment}a. The combustor was characterized in our previous work \citep{gopakumar2016mitigating,mahesh2018instability,dutta2019investigating} in a vertical configuration and is used in the present experiments in a horizontal configuration to facilitate length variation and imaging. We have achieved mitigation of thermoacoustic instability upto $\sim 20$ dB for the horizontal setup under similar operating conditions as the vertical one, reported in the earlier works \cite{dutta2019investigating,mahesh2018instability}. This establishes that the overall functioning and observations from the experimental setup remain unaffected by the change in orientation. A lean mixture ($\phi$ = 0.68) consisting of 6.5 SLPM (standard liter per minute) of methane and 90 SLPM of air (Reynolds number, $Re$ = $6 \times 10^3$ based on swirler diameter, with an uncertainty of $\pm 0.8 \%$) is supplied into the settling chamber through four equally spaced inlet ports. The equivalence ratio is calculated as $\phi = (\dot{m}_f/\dot{m}_a)_\textrm{actual}/(\dot{m}_f/\dot{m}_a)_\textrm{stoichiometry}$, where $\dot{m}_f$ and $\dot{m}_a$ are the mass flow rate of the fuel and air, respectively. 

The reactant mixture flows into the combustion chamber made of quartz, having a diameter of 46 mm and a length of 60 mm, where it is ignited. An aluminum duct of 1.5 m in length is mounted over the quartz duct which act as a resonator to generate self-excited thermoacoustic instability in the combustor. The flame stabilization is achieved using the swirler having a diameter of $30$ mm and consisting of eight straight vanes inclined at $\delta= 30^{\circ}$  with the axis is mounted on the central shaft of the motor. The geometric swirl number is obtained as $S = 2/3 \tan\delta = 0.385$ \citep{candel2014dynamics}. The swirler is actuated through a stepper motor to a maximum speed of 2100 rpm for the suppression experiments.

The airflow rate ($\dot{m}_a$) and the fuel flow rate ($\dot{m}_f$) are controlled using digital mass flow controllers (Alicat Scientific, MCR series) with a measurement uncertainty of $\pm$($0.8\%$ of reading + $0.2 \%$ of full-scale). Acoustic signatures from the combustor are acquired using a Kistler pressure transducer (sensitivity 1.84 V/bar, uncertainty $\pm $0.2 $\%$) mounted 20 mm upstream of the dump plane. The pressure signals are acquired for a duration of 5 s at a sampling frequency of 10 kHz and digitized using a National Instruments 16-bit PCI 6251 card. A high-speed CMOS camera (Photon SA5) fitted with LaVision IRO (Intensified Relay Optics) and Tamaron 150-600 mm $f$/5-6.3 telephoto lens is used to capture the unfiltered chemiluminescence images of the flame. The camera recorded 60 mm $\times$ 60 mm of the combustion chamber onto 480 pixels $\times$ 480 pixels of the sensor at a framing rate of 2 kHz while focused at the $r-\Theta$ plane at 5 mm height from the swirler exit. We recorded $10^4$ number of images at each state of the combustor operation. We simultaneously perform acoustic pressure and imaging measurements for making quantitative assessments.

Figure \ref{fig:experiment}b illustrates the variation of acoustic pressure amplitude of the first dominant mode as a function of the swirler rotation rate obtained experimentally. The error bars in the plot show the variation between the amplitudes across three experimental runs. Figure \ref{fig:experiment}c shows the change in the dominant frequency of the acoustic pressure during a transition from thermoacoustic instability to the state of suppression. The combustor remains at thermoacoustic instability for $0$ rpm with a first mode amplitude of $~153$ dB and frequency of $~76.9$ Hz. At $1800$ rpm, the amplitude reduces to about $130$ dB and frequency of $~94.2$ Hz as intermittent oscillations emerge in the combustor. Finally, the system transitions to the state of suppression at $2100$ rpm, where broadband sound replaces the dominant acoustic mode. Consequently, we observe a suppression of approximately $30$ dB on varying the swirler rotation rate from $0$ to $2100$ rpm. 

\section{\label{sec:model}Theoretical modeling and parameter estimation}

\subsection{Acoustic modeling}\addvspace{10pt} 

The governing equations for the acoustic pressure and velocity with a source term for the heat release rate after neglecting the temperature gradient and the viscous effects are written as \cite{balasubramanian2008thermoacoustic,nicoud2009zero}:
\begin{equation} \label{eq1}
\begin{split}
    \tilde{\rho}_0\frac{\partial \tilde{u}^\prime}{\partial \tilde{t}} + \frac{\partial \tilde{p}^\prime}{\partial \tilde{x}} &= 0,\\
    \frac{\partial \tilde{p}^\prime}{\partial \tilde{t}} + \gamma \tilde{p}_0 \frac{\partial \tilde{u}^\prime}{\partial \tilde{x}} &= (\gamma -1) \dot{\tilde{q}}^\prime \delta(\tilde{x} - \tilde{x}_f),
    \end{split}
\end{equation}
where $\tilde{u}^\prime$ and $\tilde{p}^\prime$ are the acoustic velocity and pressure fluctuations, $\tilde{\rho}_o$ the mean density, $\tilde{p}_o$ the mean pressure, $\gamma$ is the ratio of heat capacity, $\tilde{x}$ the axial distance in the duct, and $\tilde{t}$ the time. The heat source $\dot{\tilde{q}}^\prime$ is the unsteady heat release rate fluctuations per unit area and assumed to be compact, and the Dirac delta function $\delta$ indicates its location at $\tilde{x}$ = $\tilde{x}_f$. $\tilde{()}$ refers to dimensional quantities.  

For the present experimental configuration, the acoustic field is assumed to be one-dimensional, and the geometry is approximated to one end closed and the other end open to the atmosphere. The appropriate basis functions which satisfy the boundary conditions for the acoustic field are chosen accordingly and form the eigenmodes of the self-adjoint part of the linearized equations. Consequently, the acoustic pressure $\tilde{p}^\prime$ and velocity fluctuations $\tilde{u}^\prime$ can then be expanded  as a series of the orthogonal basis functions, satisfying the boundary conditions as follows \cite{zinn1971application}:
\begin{equation}
\begin{split}
   \tilde{p}^\prime (\tilde{x},\tilde{t}) &= \tilde{p}_0 \sum_{j = 1}^{n} \frac{\dot{\eta}_j (\tilde{t})}{\tilde{\Omega}_j} \cos(\tilde{k}_j \tilde{x}),\\ \quad  \tilde{u}^\prime (\tilde{x},\tilde{t}) &= \frac{\tilde{p}_0}{\tilde{\rho}_0 \tilde{c}_0} \sum_{j= 1}^{n} \eta_j (\tilde{t}) \sin(\tilde{k}_j \tilde{x}),  
\end{split}
\label{eq2}
\end{equation}
where the wavenumber is given by $\tilde{k}_j$ = $(2j - 1)\pi/2\tilde{L}$ and the natural frequency by $\tilde{\Omega}_j$ = $\tilde{c}_0 \tilde{k}_j$. Here, $\tilde{L}$ is the length of the duct and $\tilde{c}_0$ is the average speed of sound. The non-dimensional time-varying coefficients of the $j^\textrm{th}$ mode for the acoustic velocity $\tilde{u}^\prime$ and acoustic pressure $\tilde{p}^\prime$ are represented by $\eta_j$ and $\dot{\eta}_j$, respectively.

\subsection{Flame modeling}

The nonlinearity in the thermoacoustic system arises from a combination of the flame response to the background turbulent flow, hydrodynamic instabilities, and acoustic fluctuations \cite{lieuwen2021unsteady}. These contributions manifest in the overall heat release rate fluctuations. We assume that the contributions from the nonlinear response of the flame is encapsulated in an ensemble of non-identical phase oscillators evolving collectively under the influence of the acoustic pressure oscillations. We formulate the phase oscillators as non-identical oscillators by assigning the frequencies to each oscillator from the frequency distribution. The frequency distribution is obtained from the experimentally measured heat release rate spectrum during the state of suppression. The interaction among the phase oscillators happens through the phase ($\Phi$) and normalized amplitude ($\hat{A}$) of the acoustic pressure fluctuations. 

In particular, the interaction function taking into account the impact of the phase of the acoustics ($\Phi$) on the phase of each oscillator ($\theta_i$) is introduced as a simple periodic function: $\sin(\Phi - \theta_i)$. Moreover, the effective strength of the coupling due to the oscillator-acoustic coupling ($\tilde{K}$) and the amplitude ($\hat{A}$) of acoustic pressure fluctuations is expressed as $\tilde{K} \hat{A}$. This proportionality of the effective coupling strength on $\hat{A}$ sets up a positive feedback loop between the coupling of the phase oscillators and the amplitude of acoustic perturbation. As the population becomes more coherent, $\hat{A}$ grows and so the effective coupling $\tilde{K} \hat{A}$ increases, which tends to include even more oscillators into the synchronized group. In this manner, the thermoacoustic feedback is incorporated in the model.

In this paper, we consider the simplest case of the evolution of coupled phase oscillators, which is expressed as \citep{kuramoto1975self,ramirez2018fireflies}:
\begin{equation} \label{eq3}
   \dot{\theta}_i(\tilde{t}) = \tilde{\omega}_i + \tilde{K} \hat{A}(\tilde{t}) \sin\left[\Phi(\tilde{t})- \theta_i(\tilde{t})\right],
\end{equation}
where $\theta_i$ is the instantaneous phase of the $i^\textrm{th}$ oscillator with $i = 1,..., N$. The frequency distribution ($\tilde{\omega}_i$) of the oscillators is centered around the acoustic frequency ($\tilde{\Omega}_0$). In the above equation, the phase of individual oscillators corresponding to the phase of local heat release rate oscillations is directly coupled to the acoustic pressure, whereas the phase oscillators are equally weighted and indirectly coupled to each other via the acoustic variable. The exact form considered in Eq. \eqref{eq3} is inspired from the theory of interacting phase oscillators encountered in complex systems theory \cite{kuramoto1975self, ramirez2018fireflies,singh2022mean} and is a natural choice, as we intend to study the underlying synchronization behavior of the thermoacoustic system. 

We express the overall heat release rate fluctuations ($\dot{q}^\prime$) as a summation of the contribution from all the instantaneous phase oscillators:
\begin{equation}
\dot{\tilde{q}}^\prime = \tilde{q}_0 \sum_{i=1}^N \sin \left[\tilde{\Omega}_0 \tilde{t} + \theta_i (\tilde{t})\right],
\label{eq4b}
\end{equation}
where the term $\tilde{q}_0$ is the amplitude of heat release rate fluctuations of an individual oscillator.

\subsection{Modeling the coupling strength}

In addition to the acoustic feedback, the swirler plays a crucial role in determining the flame response \cite{candel2014dynamics}. The swirler imparts a tangential velocity to the incoming flow through its geometry and actuation. The effect of swirl due to the geometry is quantified through the angular frequency of the geometric swirl $\tilde{\Omega}_s$, while the actuation is quantified through the frequency of shaft rotation $\tilde{\Omega}_r$. These two effects together make up the characteristic frequency of the swirler: $\tilde{\Omega}_c=\tilde{\Omega}_s+\tilde{\Omega}_r$. Thus, accounting for the competition in the acoustic ($\tilde{\Omega}_0$) and swirler ($\tilde{\Omega}_c$) frequencies, the coupling strength ($\tilde{K}$) can be expressed as \cite{dutta2019investigating}:
\begin{equation}\label{eq4a}
    \tilde{K} = C \left[ \tilde{\Omega}_0 - (\tilde{\Omega}_s + \tilde{\Omega}_r) \right],
\end{equation}
where $C$ is a model constant. The angular velocity ($\tilde{\Omega}_s$) of a swirling flow is defined as $\tilde{\Omega}_s = V \sin \delta/r$ assuming solid body rotation of the fluid element \cite{dutta2019investigating}. For incoming flow velocity $V = 3$ m/s, swirler radius $r = 11$ mm and swirler blade angle $\delta = 30^\circ$, the angular velocity imparted by the static swirler is $\tilde{\Omega}_s$ = 137 rad/s. 

\subsection{Mean-field model of synchronization}

To complete the coupled flame-acoustic model, we substitute Eq. \eqref{eq4b} and \eqref{eq2} into \eqref{eq1} and project over the basis function \cite{balasubramanian2008thermoacoustic}, which results in the following second-order ordinary differential equation:
\begin{equation}
   \ddot{\eta}(\tilde{t})  + \tilde{\alpha} \dot{\eta} (\tilde{t}) + \tilde{\Omega}_0^2  \eta (\tilde{t}) = \tilde{\beta} \tilde{\Omega}_0  \cos(\tilde{k} \tilde{x}_f) \sum_{i=1}^N \sin \left[\tilde{\Omega}_0 \tilde{t} + \theta_i (\tilde{t})\right],
\label{eq5}
\end{equation}
where $\tilde{\beta} = 2(\gamma - 1) \tilde{q}_0 / \tilde{L} p_0$. Following Matveev and Culick\cite{matveev2003model}, the term $\tilde{\alpha} \dot{\eta}$ in Eq. \eqref{eq5} is introduced to account for the acoustic damping where $\tilde{\alpha}$ is the damping coefficient. In Eq. \eqref{eq5}, we restrict our analysis to a single mode for accurately reproducing the dynamics of the overall system \citep{culick2006unsteady}.

We non-dimensionalize Eq. \eqref{eq5} and Eq. \eqref{eq3} as follows: $x = \tilde{x}/\tilde{L}$; $t = \tilde{t}\tilde{\Omega}_0 $; $k =  \tilde{k}\tilde{L}$; $\alpha = \tilde{\alpha}/\tilde{\Omega}_0$; $\beta = \tilde{\beta}/\tilde{\Omega}_0 $; $\omega_i = \tilde{\omega}_i/\tilde{\Omega}_0$; $K = \tilde{K}/\tilde{\Omega}_0$; $\Omega_s = \tilde{\Omega}_s/\tilde{\Omega}_0$; $\Omega_r = \tilde{\Omega}_r/\tilde{\Omega}_0$. Further, to aid comparison between the model and experiments, we normalize the non-dimensionalized equations using the expression: $A_\textrm{LCO}$ = $\beta \cos(k x_f) N/ \alpha$, where $A_\textrm{LCO}$ is the amplitude of acoustic pressure during limit cycle oscillations (see Appendix \ref{appA} and also refer \citep{singh2022mean}). Thus, we obtain the following set of normalized and non-dimensionalized equations:
\begin{align} \label{eq6}
\frac{d \hat{\eta}_j(t)}{d t} &= \dot{\hat{\eta}}_j(t),\notag\\
\frac{d \dot{\hat{\eta}}_j(t)}{d t} &= \frac{\alpha}{N} \sum_{i = 1}^N \sin\left[t + \theta_i(t) \right]-\alpha \dot{\hat{\eta}}(t) - \hat{\eta}(t),\\
\frac{d \theta_i(t)}{d t} &= \omega_i + K \left[ \dot{\hat{\eta}}(t) \cos\left(t + \theta_i(t)\right)+ \hat{\eta}(t)\sin\left(t + \theta_i(t) \right) \right],\notag
\end{align}
where $\hat{\eta}(t) = \eta(t)/A_\textrm{LCO}$. The terms $\hat{A}$ and $\Phi$ in Eq. \eqref{eq3} are rewritten in terms of the normalized acoustic variables ($\hat{\eta}$ and $\dot{\hat{\eta}}$) using Eq. \eqref{eq1s} and Eq. \eqref{eq3s} (see Appendix \ref{appA}).

\subsection{Optimization of model parameters}

\begin{figure}[t]
    \centering
    \includegraphics[width=0.9\linewidth]{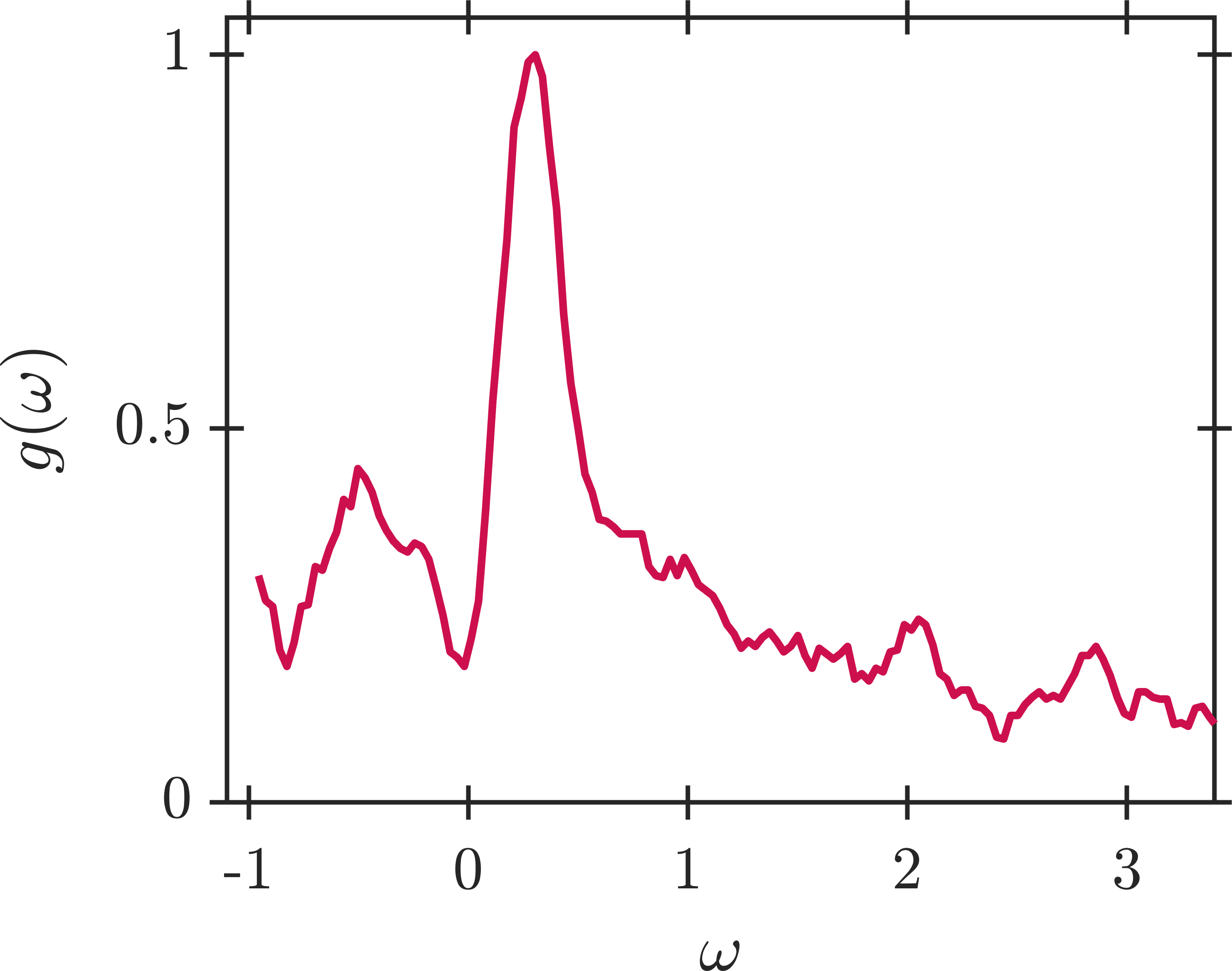}
    \caption{Oscillator frequency distribution $g(\omega)$ as a function of $\omega$ obtained from the heat release rate spectrum during a state of suppression. $\omega$ is the normalized frequency centered around the acoustic frequency.}
    \label{fig:omega_distribution}
\end{figure}

Estimating model parameters from experiments is quite challenging. Different techniques such as system identification and uncertainty quantification has been used extensively in the thermoacoustics literature \cite{polifke2010system, lee2020input, lee2021system,yu2019data,yu2021data}. In our study, we implement an optimization algorithm that minimizes the error between the experimentally obtained dynamics and the modeling results. We choose the initial phase distribution as $\theta_i(0) = \theta_m + \epsilon$, where $\theta_m$ is the mean of the phase distribution and $\epsilon$ is represented as $\epsilon \backsim \mathcal{N}(0,\sigma^2)$. The frequency distribution $g(\omega)$ of the oscillators is obtained by computing the Fourier transform of the time series of the global heat release rate fluctuations during the state of suppression shown in figure \ref{fig:omega_distribution}. The numerical procedure for sampling frequency distribution is given in Appendix of our previous work \citep{singh2022mean}. Here, we are interested in estimating the model parameters and the initial conditions, expressed as a parameter vector $\mathcal{P}$ = $[\alpha$, $K$, $\eta(0)$, $\dot{\eta}(0)$, $\theta_m$, $\sigma]$. We optimize $\mathcal{P}$ with the intention of matching the features of the model output with the experimental observations. We construct a vector for the experimental data as $\textbf{Y}_\textrm{exp}$= $[p^\prime_\textrm{exp}$, $\dot{q}^\prime_\textrm{exp}]^T$, where $p^\prime_\textrm{exp}$ and $\dot{q}^\prime_\textrm{exp}$ correspond to the normalized acoustic pressure and heat release rate oscillations, respectively. We next consider a nonlinear system using Eq. \eqref{eq6}:
\begin{equation} \label{eq26}
\dot{\textbf{Z}} = f(\mathcal{P},\textbf{Z}),
\end{equation}
where $\textbf{Z}$ = $[\dot{\hat{\eta}}$, $\hat{\eta}$, $\theta_i]^T$. Considering $\textbf{Z}_0$ as the state of a system at $t_0$, the state of a system at time $t (= t_0 + n \Delta t)$ can be estimated as:
\begin{equation} \label{eq27}
\textbf{Z}(\mathcal{P},t) = \int_{t_0}^{t} f(\mathcal{P},\textbf{Z})dt + \textbf{Z}_0.
\end{equation}
Integration in time is performed using the fourth-order Runge-Kutta scheme \citep{zheng2017modeling} with a time step of $\Delta t$. Using Eq. \eqref{eq2} and \eqref{eq4b} after appropriate normalization, we obtain:
\begin{equation} \label{eq28}
p^\prime_\textrm{model} = \dot{{\eta}}(t); \quad \dot{q}^\prime_\textrm{model} = \frac{1}{N} \sum_{i=1}^N \sin\left[t + \theta_i(t)\right].
\end{equation}
Thus, we obtain the output from the model as: $\textbf{Y}_\textrm{model}= \left[p^\prime_\textrm{model},  \dot{q}^\prime_\textrm{model}\right]$. The mean squared error (MSE) is used to construct the loss function ($\altmathcal{L}$) based on $\textbf{Y}_\textrm{model}$ and $\textbf{Y}_\textrm{exp}$: 
\begin{equation} \label{eq29}
\altmathcal{L}(\mathcal{P}) = \frac{1}{N} \sum_{n=1}^N \left \lVert \textbf{Y}_\textrm{model}(\mathcal{P},t_n) - \textbf{Y}_\textrm{exp}(\Omega_r,t_n) \right \rVert^2_2.
\end{equation}
   
The loss function is optimized with respect to the parameter vector $\mathcal{P}$. The gradient descend method is used to carry out the optimization \citep{boyd2004convex}:
   \begin{equation}
  \mathcal{P}_{i+1} = \mathcal{P}_i - \alpha_l \nabla_\mathcal{P} \altmathcal{L},
       \label{eq30}
   \end{equation}
where $\alpha_l = 1 \times 10^{-3}$, $\alpha_l$ is referred to as the learning rate and is the rate with which parameters are updated per unit gradient of loss functions. The automatic differentiation approach \citep{baydin2018automatic} is used to calculate the gradient of $\altmathcal{L}$ with respect to $\mathcal{P}$.

\section{\label{sec:result}Results and discussions}

We use the fourth-order Runge-Kutta method to solve Eq. \eqref{eq6}. The damping coefficient ($\alpha$) is estimated using the gradient descent method algorithm for parameter optimization during the state of suppression and is subsequently fixed for determining other states during a transition. We fix $N \approx 3 \times 10^3$ phase oscillators for which frequency distribution (see figure \ref{fig:omega_distribution}) effectively resolves the heat release rate spectrum during the state of suppression and where a change in $N$ has no effect on the simulation results. Using these inputs in the model, increasing the coupling strength ($K$) results in a transition to the state of suppression. We begin by confirming that a transition produced by the model is qualitatively similar to a transition obtained from the experiments. The relationship between the coupling strength ($K$) and the swirler rotation rate ($\Omega_r$) used in experiments is then determined by applying parameter optimization to each state observed in the experiments, which are tabulated in Appendix \ref{appB}.

Let us now compare the results from this model with the experimental observations. 

\begin{figure}[t]
    \centering
    \includegraphics[width=\linewidth]{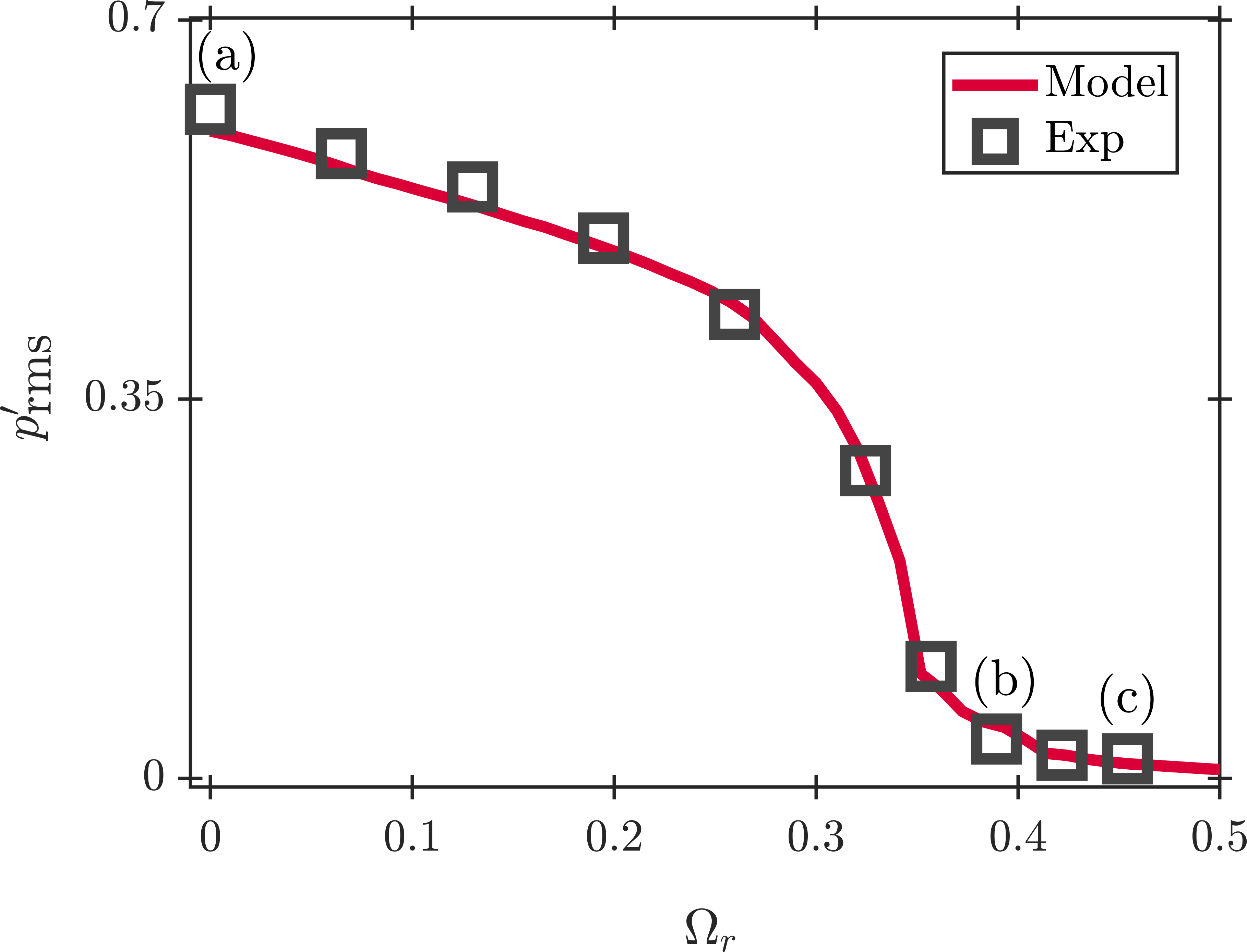}
    \caption{Comparison of the bifurcation diagrams obtained from the model (--) and experiments ($\Box$). The plot depicts the variation of the normalized $p^\prime_\textrm{rms}$ as a function of the non-dimensional swirler rotation rate ($\Omega_r$).}
    \label{fig:bifur_comp}
\end{figure}

\subsection{Bifurcation during transition to the state of suppression}

Figure \ref{fig:bifur_comp} illustrates the variation of the amplitude of the acoustic pressure ($p^\prime_\textrm{rms}$) as a function of the non-dimensional swirler rotation rate ($\Omega_r$). In this figure, we show the comparison between the bifurcation diagram obtained from the model (--) with that obtained from experiments ($\square$). To compare the transition observed in the experiments with that obtained by the model (see Eq. \eqref{eq6}), we normalize each state with the amplitude of limit cycle oscillations. Initially, when the swirler is static ($\Omega_r=0$), the combustor exhibits thermoacoustic instability with limit cycle amplitude $p^\prime_{\text{rms}} =$ 0.62, at a frequency of $f_0 = $ 76.9 Hz as shown in Fig. \ref{fig:pres_ts}a. As $\Omega_r$ increases, we notice a continuous decrease in $p^\prime_{\text{rms}}$. At the highest $\Omega_r$ value (0.455), we observe suppression, as the acoustic pressure fluctuations become low amplitude aperiodic with a broadband amplitude spectrum and low $p^\prime_\textrm{rms}$ value (see Fig. \ref{fig:pres_ts}c). 

A transition to the state of suppression occurs through the state of intermittency. This behavior can be observed at $\Omega_r$ = 0.39, where bursts of the periodic pressure oscillations appear randomly amidst low amplitude aperiodic pressure fluctuations (see Fig. \ref{fig:pres_ts}b). In figure \ref{fig:bifur_comp}, we notice that the model predicts a transition from thermoacoustic instability to the state of suppression as observed in the experiments. The monotonic decrease in $p^\prime_{\text{rms}}$ with increasing $\Omega_r$ shows that the continuous, sigmoid-type transition observed in the experiments is well captured by the model.

\subsection{Dynamical states during transition to the state of suppression}\addvspace{10pt}

\begin{figure}[t]
    \centering
    \includegraphics[width=\linewidth]{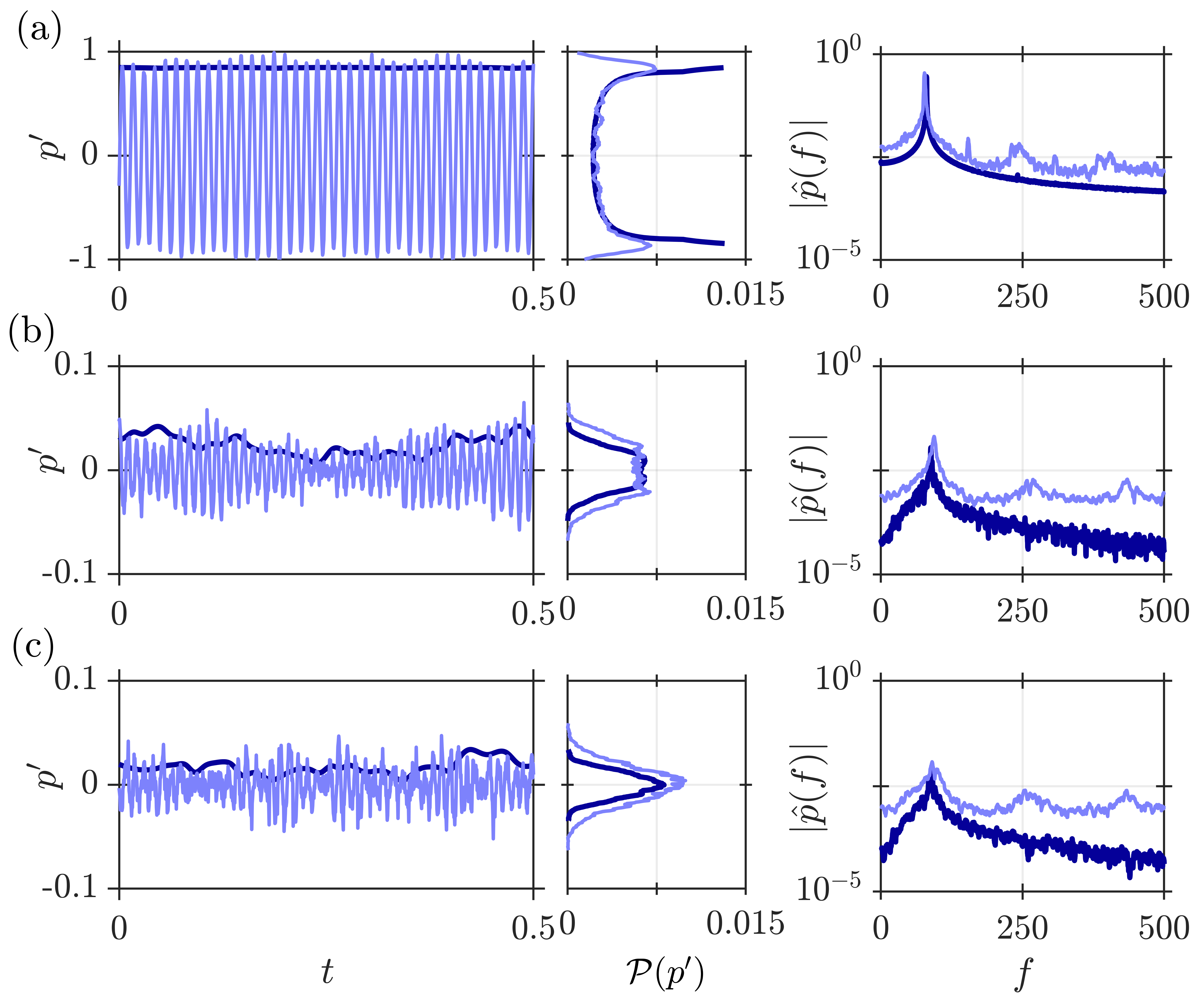}
    \caption{Comparison of the time trace, probability density function and amplitude spectrum of $p^\prime$ obtained from experiments (lighter shade) and that obtained from our model (darker shade) during the states of (a) thermoacoustic instability, (b) intermittency, and (c) suppression. (a-c) corresponds to the markers shown in Fig. \ref{fig:bifur_comp}. The envelope of time series from the model is shown in the first column for clarity.}
    \label{fig:pres_ts}
\end{figure}

\begin{figure}[h]
    \centering
    \includegraphics[width=\linewidth]{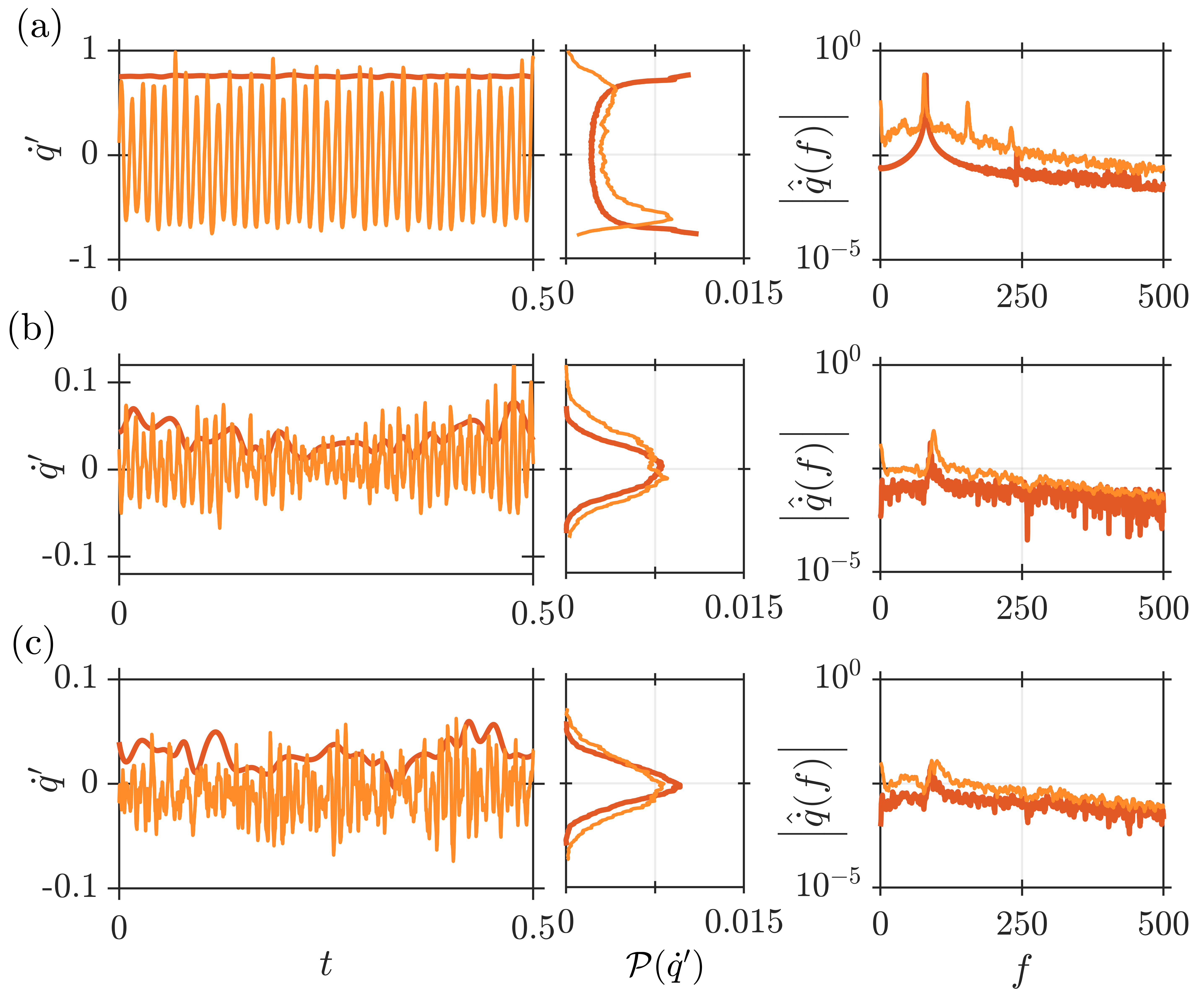}
    \caption{Comparison of the time trace, probability density function and amplitude spectrum of $\dot{q}^\prime$ obtained from experiments (lighter shade) and that obtained from our model (darker shade) during the states of (a) thermoacoustic instability, (b) intermittency, and (c) suppression. (a-c) corresponds to the markers shown in Fig. \ref{fig:bifur_comp}. Only the envelope of the time series from the model is shown.}
    \label{fig:hrr_ts}
\end{figure}

Next, we contrast the dynamics obtained from the model with the dynamical states observed during experiments at three representative states. These states correspond to thermoacoustic instability, intermittency, and suppression state at three swirler rotation rates marked (a-c) in Fig. \ref{fig:bifur_comp}. We plot the time series, probability density function, and amplitude spectrum of $p^\prime$ in Fig. \ref{fig:pres_ts} and $\dot{q}^\prime$ in Fig. \ref{fig:hrr_ts}. 

Figures \ref{fig:pres_ts}a and \ref{fig:hrr_ts}a correspond to the state of thermoacoustic instability with $\Omega_r$ = 0. We observe large-amplitude periodic oscillations in $p^\prime$ and $\dot{q}^\prime$, and the probability density function of $p^\prime$ and $\dot{q}^\prime$ is characterized by a well-defined bimodal distribution. The amplitude spectrum corresponding to the same state indicates a sharp peak at $f_0=76.9$ Hz. Figures \ref{fig:pres_ts}b \& \ref{fig:hrr_ts}b correspond to $\Omega_r$ = 0.39, and depict intermittent oscillations in both $p^\prime$ and $\dot{q}^\prime$. These intermittent oscillations lead to a change from a bimodal to a unimodal distribution in the probability density function of $p^\prime$ and $\dot{q}^\prime$. We note that the model accurately depicts the location of epochs of periodic and aperiodic oscillations in both the time series ($p^\prime$ and $\dot{q}^\prime$) while also depicting almost identical probability distribution function of both the time series. We also observe a good agreement between the amplitude spectrums of $p^\prime$ and $\dot{q}^\prime$ obtained from the experiments and the model, each showing a dominant peak at 94.2 Hz during the state of intermittency. Finally, during the state of suppression at $\Omega_r = 0.455$, we observe low-amplitude aperiodic pressure fluctuations with a well-defined unimodal probability density function (Fig. \ref{fig:pres_ts}c \& \ref{fig:hrr_ts}c). The amplitude spectrum of $p^\prime$ is broadband with a peak at $91$ Hz. Again, the match between the experiments and the model is quite evident.

Quite notably, the model yields a good match in the characteristics of the time series of the pressure and heat release rate fluctuations during various dynamical states and only requires the heat release rate spectrum during the state of suppression as an input from the experiments for obtaining $g(\omega)$. Additionally, the amplitude spectrums and probability density functions of $p^\prime$ and $\dot{q}^\prime$ for the various states of combustor operation are well approximated. Our findings demonstrate that the model accurately represents the combustor dynamics, supporting our modeling approach.

\vspace{-0.3cm}
\subsection{Relation between coupling strength and swirler rotation rate}

\begin{figure}[t]
    \centering
    \includegraphics[width= 0.9\linewidth]{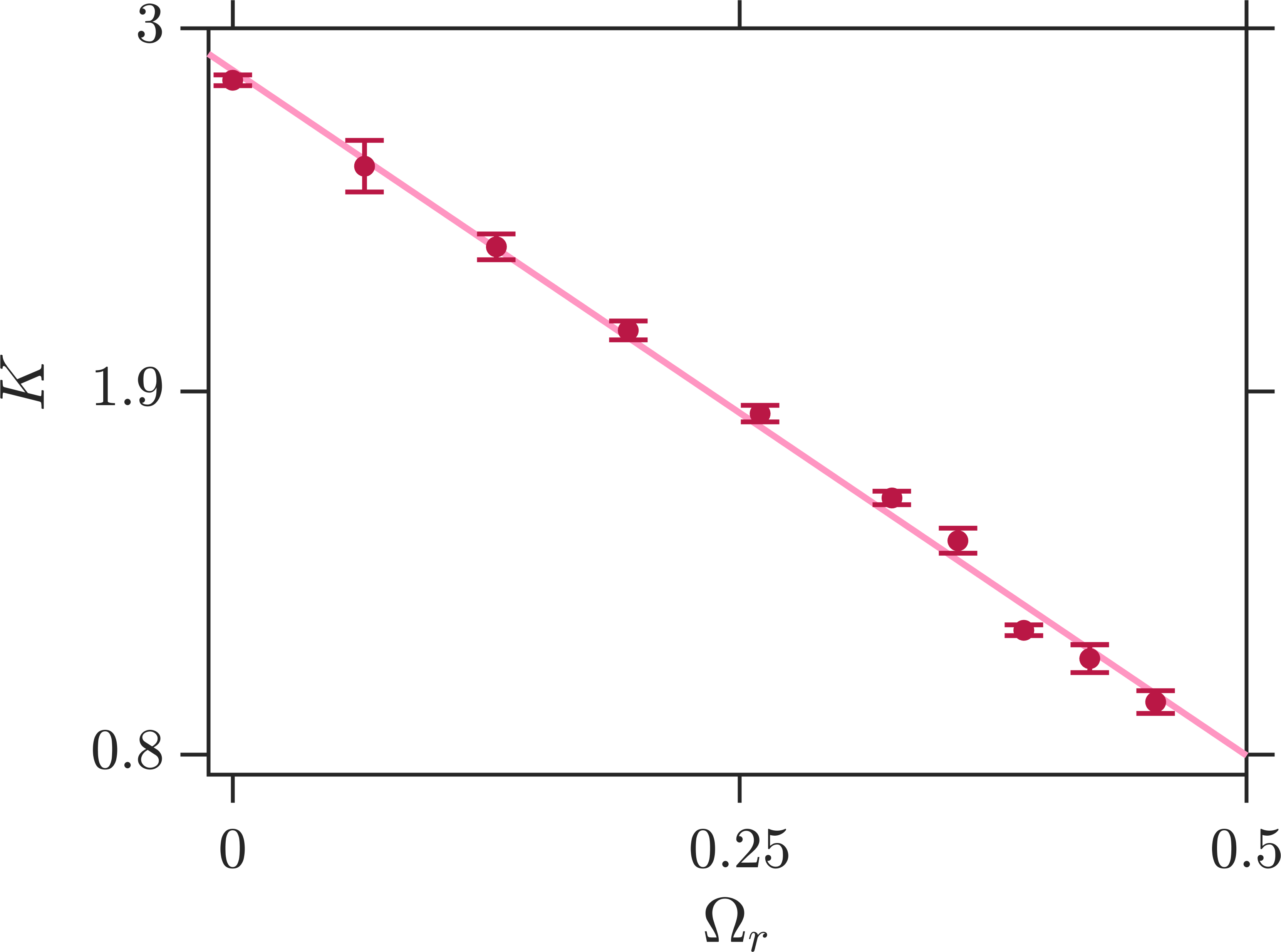}
    \caption{Mapping between the swirler rotation rate ($\Omega_r$) and coupling strength ($K$) during a transition to the state of suppression. The relation between $\Omega_r$ and $K$ is: $K = 4.1(0.71 -\Omega_r)$ with the goodness-of-fit as 0.99. The error bars represent the standard deviation in estimating $K$ by sliding the window of the time series used during optimization (Appendix \ref{appAB}).}
    \label{fig:Kvsomega}
\end{figure}

In figure \ref{fig:Kvsomega}, we show the mapping between the swirler rotation rate ($\Omega_r$) and the coupling strength ($K$), obtained by the gradient descent method using Eq. \eqref{eq30}. The error bars in the figure are determined from a distribution of $K$ for a window width of $t_\textrm{win}$ = 0.7 s using Eq. \eqref{eq30} and then sliding the window across the time series of $\textbf{Y}_\textrm{exp}$ and $\textbf{Y}_\textrm{mod}$. Please refer Appendix \ref{appAB} for a description of the window width selection process. The correspondence between the control parameters in the model and experiments will allow us to explain the experimental observations in terms of the physics embodied in the model. The estimated values strongly imply a linear relationship between the control parameter in our experiments ($\Omega_r$) and the model ($K$). The coupling strength linearly decreases according to the relation, $K = 4.1(0.71 - \Omega_r)$, thus providing \textit{a posteriori} justification for assuming a linear relationship between $K$ and $\Omega_r$ in Eq. \eqref{eq4a}. The linear relation between the $\Omega_r$ and $K$ implies that when $\Omega_r = 0$, then coupling strength among the phase oscillators is maximum, encouraging phase synchronization and leading to limit cycle oscillations. Increasing the value of $\Omega_r$ leads to a decrease in the coupling strength ($K$) among the phase oscillators, promoting phase de-synchronization and hence, the state of suppression. Thus, the model is easily interpretable in terms of experimentally relevant control parameters.

\begin{figure}[t]
    \centering
    \includegraphics[width=\linewidth]{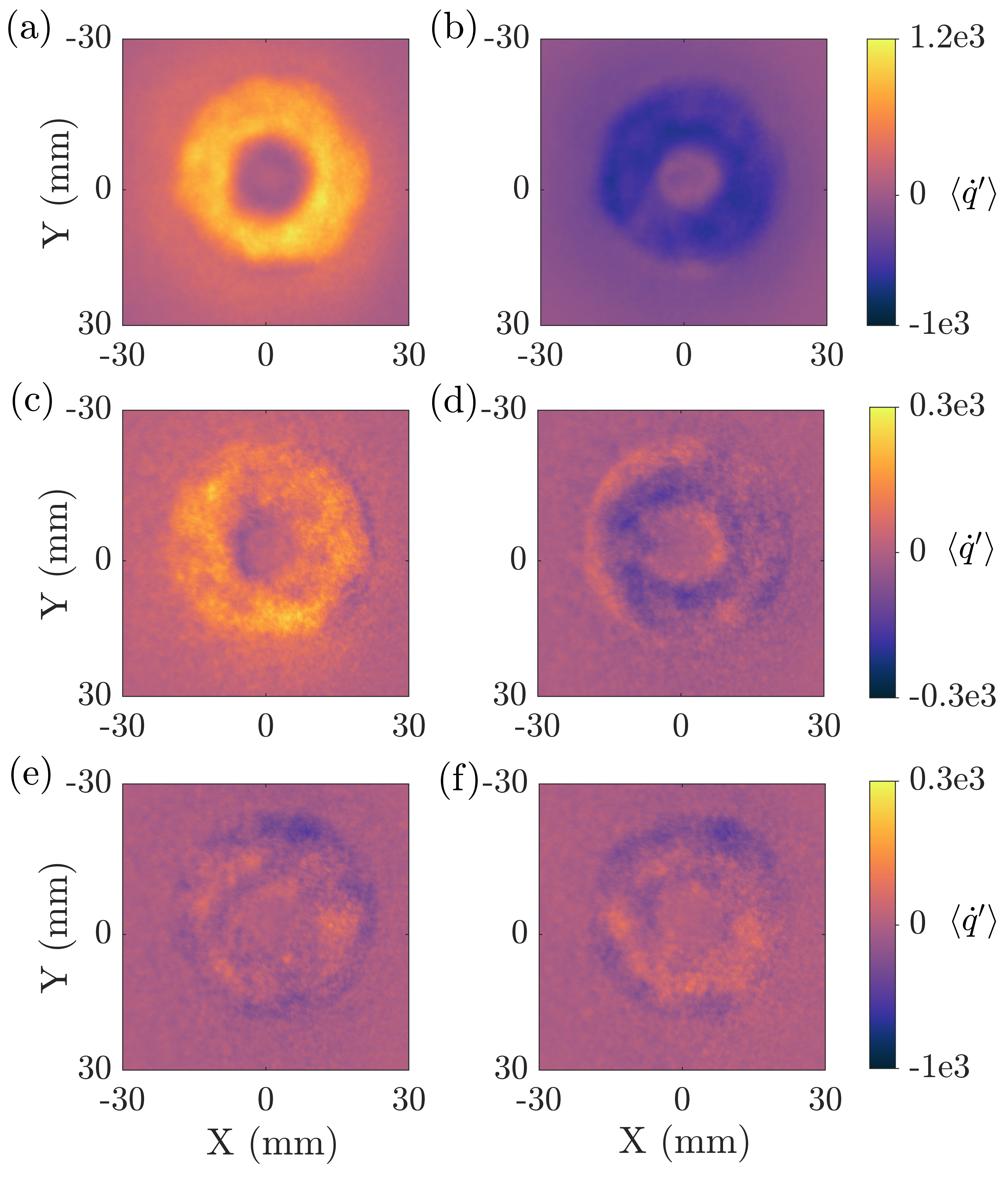}
    \caption{Mean-subtracted phase-averaged flame images at pressure maxima (left) and minima (right) during the dynamical states: (a-b) thermoacoustic instability, (c-d) periodic epochs of intermittency, and (e-f) suppression state. The flame image depicting a well-defined circular shape with very high intensity during thermoacoustic instability transitions to non-uniform flame structures with low intensity during the state of suppression.}
    \label{fig:avg_flame}
\end{figure}

\subsection{Flame dynamics during transition to the state of suppression}

\begin{figure*}[t]
    \centering
    \includegraphics[width=\linewidth]{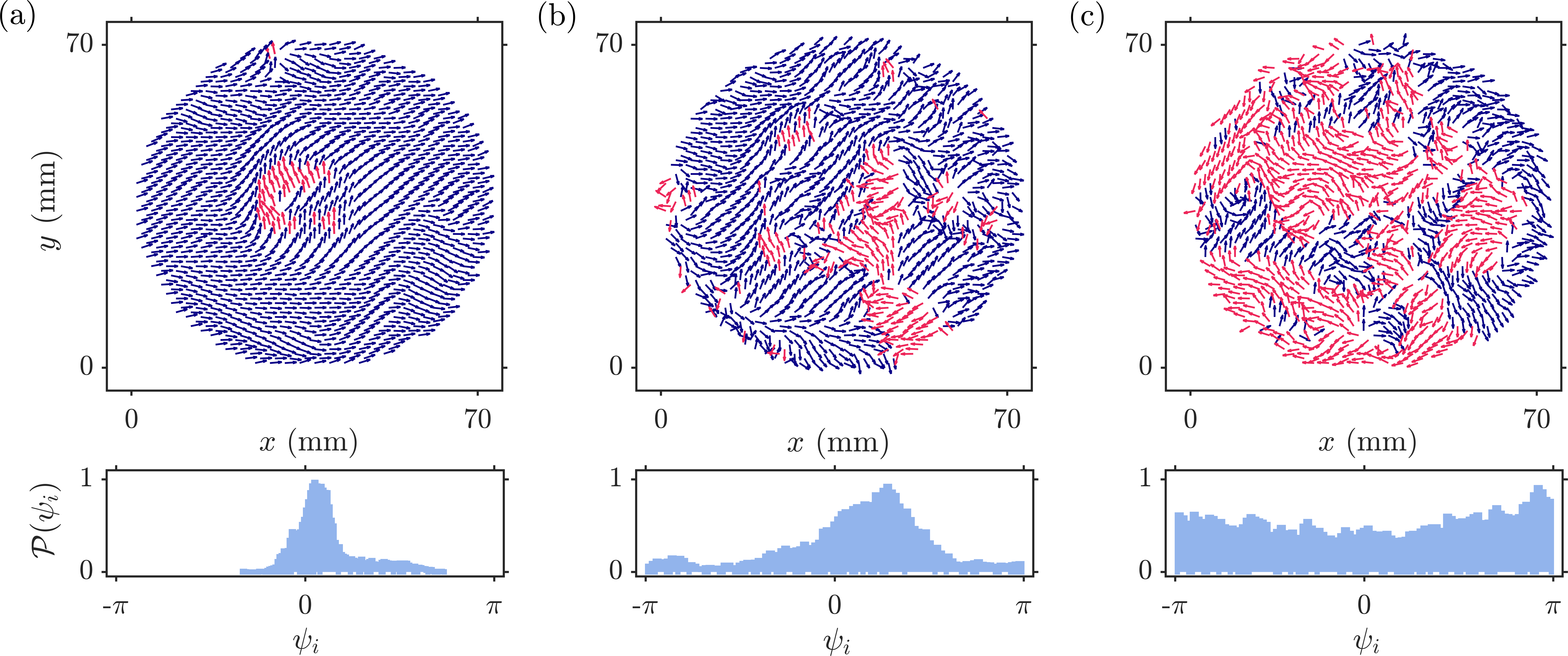}
   \caption{Disruption of order on varying the swirler rotation rate ($\Omega_r$) from 0 to 0.455 in the turbulent combustor. A typical snapshot of the spatially distributed phasors ($\psi_i$) is obtained by taking the phase difference between the local heat release rate ($\theta_i$) and acoustic oscillations ($\Phi$) during the state of (a) thermoacoustic instability, (b) intermittency, and (c) suppression. To delineate regions of acoustic power sources and sinks, phasors have been colored blue if $|\psi_i| < \pi/2$ and red otherwise. (Multimedia View).}
    \label{fig:phasor}
\end{figure*}

We now demonstrate the flame images obtained from the experiments to better understand the dynamics of the flames during a transition to the state of suppression. Figure \ref{fig:avg_flame} shows the phase-averaged unfiltered chemiluminescence images at pressure maxima (left column) and minima (right column) during the three dynamical states. The maxima and minima locations correspond to $90^{\circ}$ and $270^{\circ}$ phase, which are found from the instantaneous phase values of $p^\prime$ data using the Hilbert transform \citep{pikovsky2002synchronization}. During thermoacoustic instability, at pressure maxima (Fig. \ref{fig:avg_flame}a), the mean-subtracted flame image depicts a well-defined circular shape with very high intensity, which reaches a negative value at pressure minima (Fig. \ref{fig:avg_flame}b). Thus, the flame fluctuations are strongly correlated with the pressure fluctuations. Moreover, the flame structure shows flame stabilization along the central shear layer separating the inner and outer recirculation zones \cite{mahesh2018instability}. 

Next, we consider the flame dynamics during the periodic part of intermittency (Figs. \ref{fig:avg_flame}c-d). The phase-averaged flame image during maxima and minima still depict high and low-intensity values, respectively, albeit with lower intensity as compared to thermoacoustic instability. While the flame structure is similar to that during thermoacoustic instability, the distribution is diffused during both maxima and minima. This implies that some parts of the flame may not be attaining maxima and minima at the instant of pressure maxima and pressure minima. Thus, the flame fluctuations are weakly correlated with the pressure fluctuations. Finally, figures \ref{fig:avg_flame} (e-f) correspond to the state of suppression. The phase-averaged chemiluminescence images corresponding to the pressure peaks (Fig \ref{fig:avg_flame}e) and troughs (Fig \ref{fig:avg_flame}f) are similar to each other. We observe incoherent and non-uniform flame structures with low intensity in comparison to the periodic part of intermittency and thermoacoustic instability. Moreover, the distributions of intensity between pressure maxima and minima are virtually indistinguishable, implying no correlation between flame fluctuations and acoustic fluctuations.

We are aware that any fluctuation in the flame results in an unsteady heat release rate which, in turn, causes acoustic disturbances in the combustion chamber and affects the flame when reflected from an appropriate acoustic boundary. The strength of the periodic fluctuations in pressure and heat release rate starts increasing as the feedback between the acoustic field and heat release rate fluctuations increases. In figure \ref{fig:avg_flame}, we notice the difference in the flame intensities when the swirler rotation rate is systematically varied. As a result, by varying the swirler rotation rate, we are disrupting the strength of a feedback loop to bring the sustained high-intensity oscillations in the flame at thermoacoustic instability to the low-intensity flame fluctuations at the state of suppression.

\subsection{Synchronization transition to the state of suppression}

Furthermore, we investigate the spatio-temporal dynamics during the transition from thermoacoustic instability to the state of suppression obtained from the experiments. To that end, we analyze the characteristics of synchronization in a spatially extended thermoacoustic system by investigating the coupled behavior of the acoustic pressure $p^\prime(t)$ and the local heat release rate fluctuations $\dot{q}^\prime (x,y,t)$. The local heat release rate fluctuations are extracted from the intensity variation observed at each pixel of the time-resolved chemiluminescence images. The flame images are coarse-grained over $6 \times 6$ pixels to minimize the effect of noisy fluctuations in them. We install the pressure transducer near to the location of the flame imaging, to avoid acoustic phase delay effects in our experimental measurements. Figure \ref{fig:phasor} (Multimedia View) depicts the spatial distribution of instantaneous phase ($\psi_i$) during a transition from thermoacoustic instability to the state of suppression through the state of intermittency when the swirler rotation rate ($\Omega_r$) is increased. The instantaneous phasor field ($\psi_i$) is obtained by subtracting the phase of the acoustic pressure ($\Phi$) from the phase of the local heat release rate fluctuations ($\theta_i$). The instantaneous phase of the local heat release rate fluctuations is obtained using Hilbert transformation.

In figure \ref{fig:phasor}a (Multimedia View), when the swirler is static ($\Omega_r = 0$), the acoustic pressure and heat release rate oscillate in phase, leading to a coherent field of the phasors. In the probability density function of $\psi_i$, we notice that the phase values are mostly $\psi_i < |\pi/2|$ radians, leading to enhanced acoustic driving during the occurrence of thermoacoustic instability in the combustor and hence satisfying the Rayleigh criterion. We notice that the spatial synchrony in the phase plot starts reducing with an increase in $\Omega_r$. For instance, during the state of intermittency at $\Omega_r = 0.39$, the phase-field shows both coherent and incoherent fields of phasors shown in figure \ref{fig:phasor}b (Multimedia View). The probability density function of $\psi_i$ associated with intermittency is broadening in comparison with the probability density function of $\psi_i$ obtained during thermoacoustic instability. In this state, there is the coexistence of clusters of both spatial synchrony and asynchrony in the phase field, referred to as a chimera state \citep{sujith2020complex}. Further, increasing $\Omega_r$ to 0.48, we observe that the phase field is randomly oriented and incoherent (Fig. \ref{fig:phasor}c (Multimedia View)). In this state, the heat release rate fluctuations are dominated only by the turbulent flow, which results in a de-synchronized field of the phasors and a broadband distribution of $\altmathcal{P}(\psi_i)$. This asynchronous behavior of the local heat release rate fluctuations during the state of suppression prevents the pressure oscillations from increasing in the amplitude, which leads to low-amplitude aperiodic fluctuations in the temporal dynamics of both acoustic pressure ($p^\prime$) and global heat release rate fluctuations ($\dot{q}^\prime$).

\begin{figure*}[t]
    \centering
    \includegraphics[width= 1.001\linewidth]{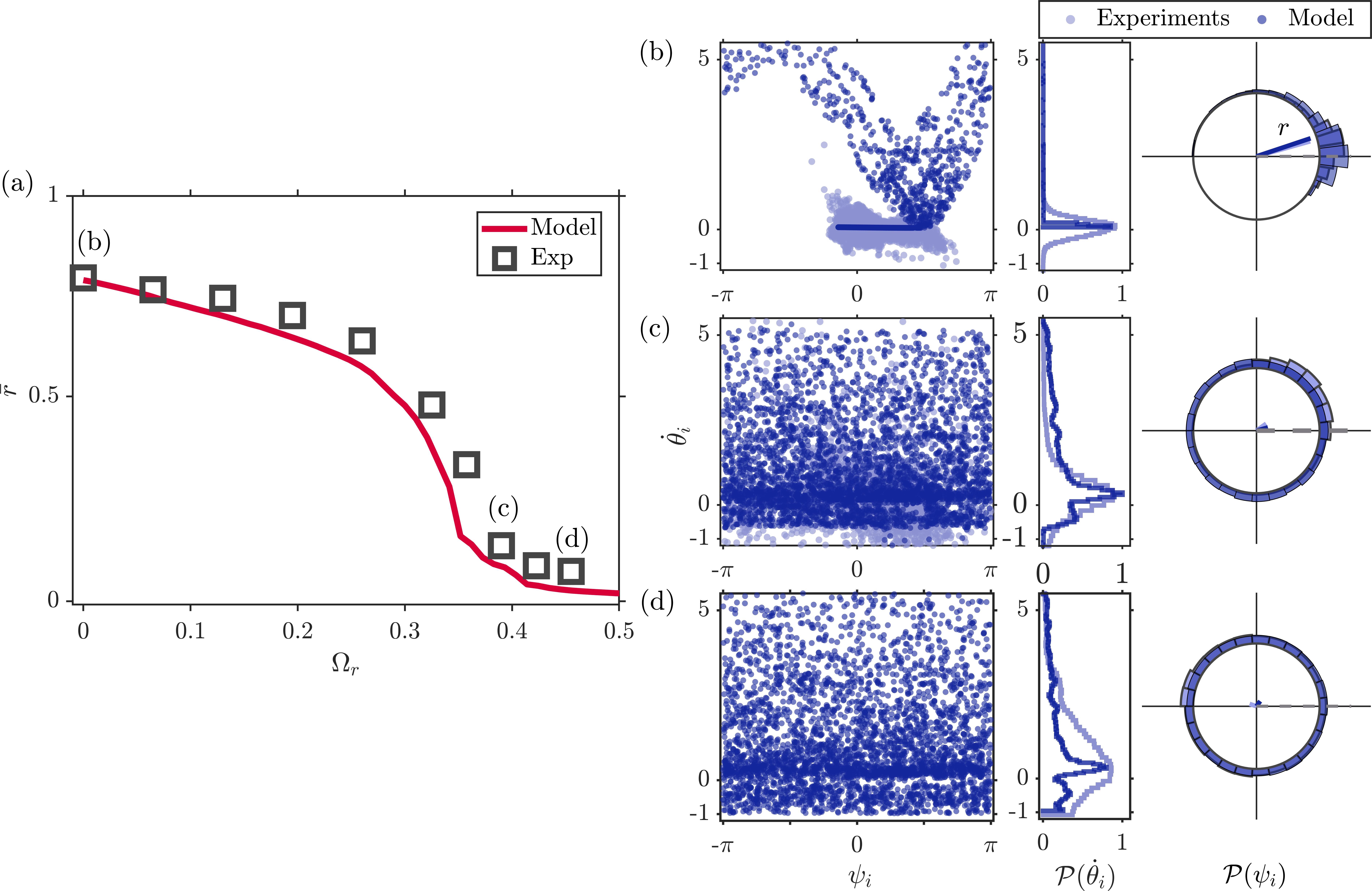}
    \caption{(a) The synchronization bifurcation diagram depicts the variation of time-averaged order parameter ($\bar{r}$) as a function of non-dimensional swirler rotation rate ($\Omega_r$). The typical instantaneous oscillator distribution in the $\dot{\theta}_i - \psi_i$ plane, as well as the distribution of $\altmathcal{P}(\dot{\theta}_i)$ and $\altmathcal{P}(\psi_i)$ during the state of (b) thermoacoustic instability, (c) intermittency, and (d) suppression for experimental (lighter marker) and modeling (darker marker) dataset. The Kuramoto order parameter ($r$) from the experiments and model (--) is shown in the last column. (Multimedia View)}
    \label{fig:order_param}
\end{figure*}

We now quantify the characteristics of synchronization through a measure called the Kuramoto order parameter \citep{strogatz2000kuramoto}. The time-averaged order parameter ($\bar{r}$) is defined as:
\begin{equation} \label{eq31}
\bar{r} = \frac{1}{N}\bigg\langle \bigg|\sum_{i=1}^N \exp\left(i\theta_i(t)\right)\bigg|\bigg\rangle_t,
\end{equation}
where $\theta_i$ is the phase of the $i^\textrm{th}$ phase oscillator and $\langle \cdot\rangle_t$ implies time average. The order parameter is defined as the degree of synchrony among the oscillators and it varies between $[0,1]$. A value of $\bar{r}$ close to zero indicates de-synchronized states, whereas a value of $\bar{r}$ close to one indicates synchronized states.

Figure \ref{fig:order_param}a depicts the variation of the order parameter as a function of the non-dimensional swirler rotation rate ($\Omega_r$). The time-averaged order parameter ($\bar{r}$) from the model is determined using Eq. \eqref{eq31}, while $\bar{r}$ from the experiments is determined according to Eq. \eqref{eq:appC1} (see Appendix \ref{appC}). The gradual decrease in the order parameter ($\bar{r}$) indicates a transition from an ordered state where the heat release rate oscillators are in synchrony to a disordered state where the oscillators are in asynchrony. 

To compare the synchronization observed in the spatial field of the experiments (Fig. \ref{fig:phasor}) and from the model (Eq. \eqref{eq6}), we plot the characteristics of the oscillators in the $\dot{\theta}_i - \psi_i$ phase space. Here, $\dot{\theta}_i$ is the instantaneous frequency and $\psi_i$ is the phase difference between $\theta_i$ and $\Phi$. Figure \ref{fig:order_param}(b-d) (Multimedia View) shows the instantaneous oscillator distribution on the $\dot{\theta}_i - \psi_i$ plane during different dynamical states (first column). These plots also include the distribution of instantaneous frequency $\altmathcal{P}(\dot{\theta}_i)$ in the middle column and the distribution of the instantaneous relative phase $\altmathcal{P}(\psi_i)$ along with the order parameter ($r$) in the last column. The distribution of $\psi_i$ is shown in polar coordinates (last column), and the frame of reference of the oscillators is co-rotating with the frequency of the acoustic pressure ($\Omega_0$). The oscillators obtained from the spatio-temporal measurements are shown in lighter shades of marker, while those obtained from the model are shown in darker shades of marker. Since the heat release rate fluctuations from the experiments are spatially distributed, comparing the characteristics of oscillators in the $\dot{\theta}_i-\psi_i$ plane allows us to evaluate how well the low-dimensional mean-field synchronization model captures the characteristics of the spatio-temporal synchronization.

In figure \ref{fig:order_param}(b) (Multimedia View), when the swirler is static ($\Omega_r = 0$) corresponding to thermoacoustic instability (see Fig. \ref{fig:pres_ts}a), we observe that the oscillators are entrained at the acoustic frequency and are phase-locked with the distribution of $\psi_i$ mostly between $-\pi/2 $ and $\pi/2$ radians (first column). In the last two columns, we notice a sharp peak, narrowband distribution of frequency $\altmathcal{P}(\dot{\theta}_i)$ and relative phase $\altmathcal{P}(\psi_i)$. The value of order parameter $\bar{r}$ is 0.8, implying global phase synchronization among the phase oscillators. In figure \ref{fig:order_param}(c) (Multimedia View), when the swirler rotation rate is $\Omega_r = 0.39$ corresponding to the state of intermittency (see Fig. \ref{fig:pres_ts}b), we notice the larger regions of phase-synchronized clusters where $\psi_i < |\pi/2|$ at some spatial locations and $\psi_i > |\pi/2|$ at other locations, and comparatively less narrowband distribution of $\altmathcal{P}(\dot{\theta}_i)$ and $\altmathcal{P}(\psi_i)$. The order parameter is $\bar{r} = 0.13$ during the state of intermittency. Finally, in figure \ref{fig:order_param}d (Multimedia View), when the swirler rotation rate is $\Omega_r = 0.455$ corresponding to the state of suppression (see Fig. \ref{fig:pres_ts}c), the oscillators are distributed in $\psi_i > |\pi/2|$, implying the phase de-synchronized among the oscillators. We also observe a broadband distribution in the distribution of $\altmathcal{P}(\dot{\theta}_i)$ and $\altmathcal{P}(\psi_i)$ and a value of $\bar{r}$ close to zero. Thus, the gradual disappearance of the order among the oscillators, which is associated with the continuous shrinking of the size of the cluster of oscillators, leads to a continuous de-synchronized transition.

Although the model does well in capturing the bifurcation characteristics and aspects of synchronization, it does not capture the higher modes of the spectrum (see Figs. \ref{fig:pres_ts} and \ref{fig:hrr_ts}). This is by construction of the model as we did not consider higher modes in Eq. \eqref{eq2}, to keep the analysis simple. Additionally, the prediction of dynamical states on further increase in swirler rotation rate beyond the state of suppression from the model is left for future studies. 

\section{\label{sec:conclu}Conclusion}

In this paper, we report experiments and modeling of open-loop control of thermoacoustic instability in turbulent combustion system using an actuated swirler. A systematic increase in the rotation rate of the swirler leads to the suppression of limit cycle oscillations as the dynamics of the combustor transitions to low-amplitude aperiodic oscillations through an intermediate state of intermittency. We propose a mean-field synchronization model for the flame response comprising an ensemble of non-identical phase oscillators evolving collectively under the effect of acoustic pressure. The effect of the active swirler is incorporated naturally into the model in terms of the relative time scales of swirler rotation and acoustic frequency. We further implement a parameter identification technique to obtain exact correspondence between the model parameter values and experimental observations. We find that the mapping between the swirler rotation rate in the experiments and the coupling strength in the model manifests as a linear relationship between them. 

Through a comparison of the bifurcation diagram, time series, probability density functions and amplitude spectrums, we show that the model replicates the experimentally observed $p^\prime$ and $\dot{q}^\prime$ fluctuations very well. Further, we show that the model captures the characteristics of spatio-temporal synchronization underlying a transition to the state of suppression while depicting states such as synchronization, chimera, and de-synchronization. In particular, we find that the phase oscillators are synchronized during thermoacoustic instability, partially synchronized during intermittency, and undergo progressive de-synchronization during the suppression. Therefore, we notice a sigmoid-type transition to the suppression state happens through the underlying synchronization. As a consequence, using the mean-field thermoacoustic model, we establish that the active swirler suppresses thermoacoustic oscillations through a de-synchronization transition.

\begin{acknowledgments}
S. Singh, A. K. Dutta, and J. M. Dhadphale gratefully acknowledge the Ministry of Human Resource Development (MHRD) for Ph.D. funding through the Half-Time Research Assistantship (HTRA). R. I. Sujith is grateful for the funding from the IoE initiative of IIT Madras (SB/2021/0845/AE/MHRD/002696) and the Office of Naval Research (ONR) Global (contractor monitor: Dr. R. Kolar; Grant No. N62909-18-1). A. Roy acknowledges the Post Doctoral Researcher Fellowship under the same IoE grant. S. Chaudhuri acknowledges support from Ontario Research Fund, and from Natural Sciences and Engineering Research Council of Canada through the Discovery Grant (RGPIN-2021-02676). 
\end{acknowledgments}

\section*{Data Availability Statement}

The data that support the findings of this study are available from the corresponding author upon reasonable request.

\appendix

\section{\label{appA}Calculating $A_\textrm{LCO}$ using the method of averaging} 

To perform the comparison between the experimental and modeling results, we need to normalize the amplitude of acoustic pressure during various states using the amplitude of acoustic pressure at the limit cycle oscillations (LCO). We use the method of averaging \cite{balanov2008synchronization} in the model to determine the amplitude of acoustic pressure during LCO ($A_\textrm{LCO}$). Using the method of averaging, Eq. \eqref{eq5} in the time-domain describes the evolution of pressure fluctuations that can be mapped onto a pair of coupled ordinary differential equations by introducing two slow flow variables: amplitude $A$ and phase $\Phi$. This decomposition is commonly known as \textit{Kryloff-Bogolyubov} decomposition \citep{krylov1947introduction}. We start by decomposing $\eta (t)$ as:
	\begin{equation}
	\eta (t) = - A(t)\cos\left(t + \Phi (t)\right),
	\label{eq1s}
	\end{equation}
	where the amplitude $A(t)$ represents the envelope of the acoustic pressure and the phase $\Phi (t)$ represents the time-dependent phase shift. Both $A(t)$ and $\Phi (t)$ are slowly varying quantities because they change on a longer timescale than the period of oscillations of the thermoacoustic mode of interest. Next, we differentiate $\eta(t)$ with respect to $t$:
	
	\begin{equation} \label{eq2s}
     \begin{aligned}
	\dot{\eta} (t) = &-\dot{A}(t)\cos\left(t + \Phi (t)\right) + A(t) \sin\left(t + \Phi (t)\right) \\ &+ A(t)\dot{\Phi} (t) \sin\left(t + \Phi (t)\right). 
    \end{aligned}
    \end{equation}

     On expressing $\eta(t)$ as a function of $A(t)$ and $\Phi(t)$ in Eq. \eqref{eq1s}, we introduced additional ambiguity in the equation which can be removed by prescribing an arbitrary relationship between these quantities as: $- \dot{A}(t) \cos\left(t + \Phi (t)\right) + A(t) \dot{\Phi} (t) \sin\left(t + \Phi (t)\right) = 0$, which in turn leads to a simple expression for $\dot{\eta} (t)$ of the form:
	\begin{equation}
	\dot{\eta} (t) =  A(t) \sin\left(t + \Phi (t)\right).
	\label{eq3s}
	\end{equation}
	Next, we use Euler's formula to represent trigonometric functions present in $\eta$ and $\dot{\eta}$, which leads to:
	\begin{equation}
	\begin{aligned}
	\eta (t) &= - \frac{1}{2 } \big(A(t) e^{i \Phi (t)} e^{it}  + A(t) e^{-i \Phi (t)} e^{-it}  \big) \\ &=  - \frac{1}{2}\big( a e^{it} + a^\ast e^{-it} \big),
	\label{eq4s}
	\end{aligned}
	\end{equation}
	and
	\begin{equation}
	\dot{\eta} (t) = A(t) \frac{e^{i(t + \Phi (t))} - e^{-i(t + \Phi (t))}}{2i} = - \frac{i}{2} \big(a e^{i t} - a^\ast e^{-i t} \big).
	\label{eq5s}
	\end{equation}
	Here, $a$ is a complex function of time, such that $a = A(t) e^{i \Phi (t)}$ and $a^\ast = A(t) e^{- i \Phi(t)}$, where the asterisk denotes the complex conjugate. We also calculate $\ddot{\eta}(t)$ by differentiating $\dot{\eta} (t)$ w.r.t $t$ and we get:
	\begin{equation}
	\ddot{\eta} (t) = -\frac{i}{2} \big( \dot{a}e^{i t} - \dot{a}^\ast e^{-i t} \big) + \frac{1}{2} \big(a e^{i t} + a^\ast e^{-i t}\big).
	\label{eq6s}
	\end{equation}
	We then substitute $\eta(t)$, $\dot{\eta}(t)$ and $\ddot{\eta}(t)$ ( Eqs. \eqref{eq4s}, \eqref{eq5s} and \eqref{eq6s}, respectively) into Eq. \eqref{eq5} and obtain:
	\begin{equation}
	\begin{split}
	& -i\big( \dot{a}e^{i t}  - \dot{a}^\ast e^{-i t} \big) +  \big(a e^{i t} + a^\ast e^{-i t}\big)   - i \alpha \big(a e^{i t} - a^\ast e^{-i t} \big) \\ & - \big( a e^{it} + a^\ast e^{-it} \big)  = -i\beta \cos(kz_f) \sum_{i=1}^N \big(e^{i(t + \theta_i (t))} - e^{-i(t + \theta_i (t))}\big).
	\label{eq7s}
	\end{split}
	\end{equation}
	By canceling the second and fourth terms in the above equation and then multiplying the whole equation by $e^{-i t}$, we obtain:
	\begin{equation}
	\begin{split}
	&\big(\dot{a}  - \dot{a}^\ast e^{-2i t} \big)  +  \alpha \big(a - a^\ast e^{-2 i t} \big) \\ &=  \beta \cos(kz_f) \sum_{i=1}^N \big(e^{i\theta_i(t)} - e^{-2it} e^{-i \theta_i (t))}\big).
	\label{eq8s}
	\end{split}
	\end{equation}
	Note that $a$, $\dot{a}$ and their complex conjugate are slow functions of time as compared to the functions $e^{\pm n i t}$, where $n$ is an integer. This means that the slow flow variables do not change much during one period of fast oscillations. If we average the whole equation over one period of fast oscillations, i.e., $T = 2 \pi$, we can get rid of the terms corresponding to the fast time scale, and only the terms related to the slow time scale will remain in the equation. The time average $\bar{f}$ of a smooth function $f(t)$ over the time interval $T$ is defined as $\bar{f} = \frac{1}{T} \int_0^{2 \pi} f(t) dt$. It is easy to see that all the terms containing $e^{-2i t}$ would integrate to zero over the time oscillation period $T$. Thus, on applying time averaging on Eq. \eqref{eq8s}, we obtain:
	\begin{equation}
	\dot{a} + \alpha a = \beta \cos(kz_f) \sum_{i=1}^N e^{i\theta_i(t)}.
	\label{eq9s}
	\end{equation}
	Recalling that $a = A(t) e^{i \Phi (t)}$ and substituting it in the above equation, we get:
	\begin{equation}
	\dot{A}(t) e^{i \Phi(t)} + i A(t) \dot{\Phi}(t) e^{i\Phi(t)} + \alpha A(t) e^{i \Phi (t)} =  \beta \cos(kz_f) \sum_{i=1}^N e^{i\theta_i(t)}.
	\label{eq10s}
	\end{equation}
	On multiplying Eq. \eqref{eq10s} by $e^{-i \Phi(t)}$ followed by rewriting the exponential term back in the trigonometric quantities and finally, splitting the equation into real and imaginary parts yields the following equations:
	\begin{subequations} 	\label{eq11s}
		\begin{align}
		\dot{A}(t) &= \beta \cos(kz_f) \sum_{i=1}^N \cos(\theta_i(t) - \Phi(t)) - \alpha A(t), \label{eq11sa} \\ 	
		\dot{\Phi}(t) & = \frac{\beta}{A(t)} \cos(kz_f) \sum_{i=1}^N \sin(\theta_i(t) - \Phi(t)). \label{eq11sb}
		\end{align}
	\end{subequations}

	During limit cycle oscillations, we consider $\dot{A}(t) = 0$ and the phase difference between the acoustic pressure and heat release rate is almost zero. Therefore, on considering $\theta_i (t) - \Phi(t) \approx$ 0 in Eq. \eqref{eq11sa} turns out as: 
	\begin{equation} \label{eq12s}
	 A_\textrm{LCO} = \beta \cos(k z_f) N / \alpha.
	\end{equation}
This expression of the amplitude of acoustic pressure during thermoacoustic instability is then used for normalizing Eq. \eqref{eq5}.

\section{\label{appAB} Parameter sensitivity analysis}

\begin{figure}[t]
    \centering
    \includegraphics[width=0.85\linewidth]{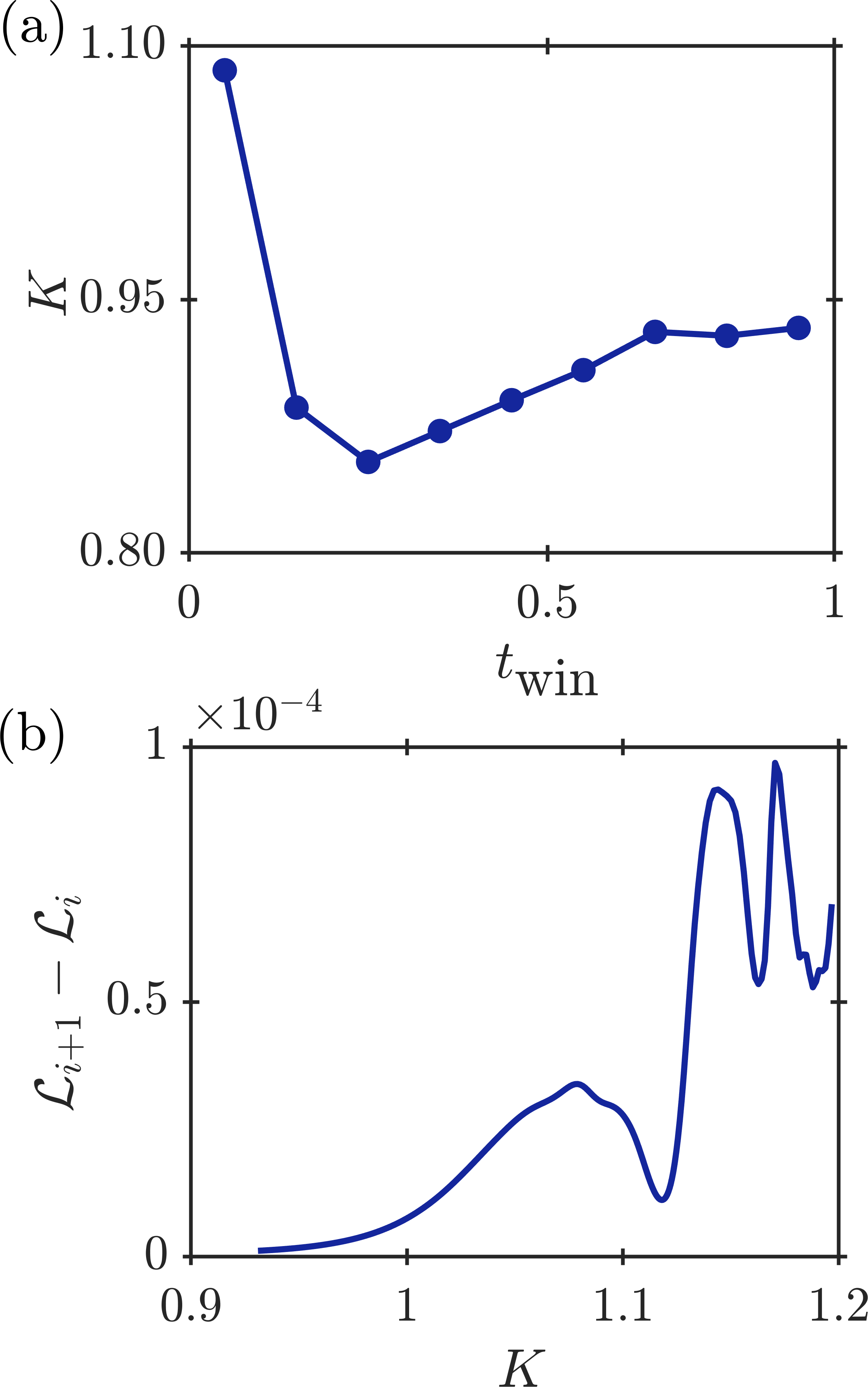}
    \caption{Panel (a) depicts the convergence of the coupling strength ($K$) as a function of the time window width ($t_\textrm{win}$) of the time series of $\textbf{Y}_{\textrm{model}}$ and $\textbf{Y}_{\textrm{exp}}$ over 500 iterations using the parameter optimization algorithm. The initial value of $K$ used is 1.2 and $t_\textrm{win}$ is varied from 0.05 to 1 s. Panel (b) shows optimization of $K$ by minimizing the loss function ($\altmathcal{L}_i$) over 500 iterations by taking difference between $\altmathcal{L}_{i+1}$ and $\altmathcal{L}_{i}$. The sensitivity analysis on parameter estimation shown here is carried out during the state of suppression.}
    \label{fig:sensitivty}
\end{figure}

In order to estimate the model parameters and the initial condition, expressed as $\mathcal{P}$, we minimize the error between $\textbf{Y}_{\textrm{model}}$ and $\textbf{Y}_{\textrm{exp}}$. We take the portion of the time series of both $\textbf{Y}_{\textrm{model}}$ and $\textbf{Y}_{\textrm{exp}}$ for estimating $\mathcal{P}$. We first find out what portion of the time series of $\textbf{Y}_{\textrm{model}}$ and $\textbf{Y}_{\textrm{exp}}$ is enough for convergence of the model parameters. The window width of the time series $t_\textrm{win}$ is varied from 0.05 to 1 s for obtaining $\mathcal{P}$ over 500 iterations. In figure \ref{fig:sensitivty}a, we show only the convergence of the coupling strength ($K$) as a function of $t_\textrm{win}$ during the suppression state. We observe that the value of $K$ remains constant after $t_\textrm{win} = 0.6$ s. Thus, we use $t_\textrm{win} = 0.7$ s for estimating parameters across all the datasets.

The width of the time window is then fixed, and the window is moved across the entire time series of $\textbf{Y}_{\textrm{model}}$ and $\textbf{Y}_{\textrm{exp}}$ to determine the range of $K$ variation at each dynamical state. Figure \ref{fig:sensitivty}b shows the realization of a minimization scheme used to obtain optimized $K$. The plot shows the reduction of the difference in $\altmathcal{L}_{i+1}$ and $\altmathcal{L}_{i}$ with change in iterations started with an initial guess of $K = 1.2$. Hence, the estimated parameter used in our paper are properly optimized and verified using the above technique.

\section{\label{appB} Estimated parameter values during the transition}

The estimated model parameters during thermoacoustic instability, intermittency and suppression state are shown in table \ref{tab:table1}. The damping coefficient ($\alpha$) is obtained by the gradient descent method using Eq. \eqref{eq30} during the state of suppression and has a value of 0.27.

\begin{table}[h]
    \caption{Estimated parameter values for the different dynamical states marked as a-c in Fig. \ref{fig:bifur_comp}.}
\centerline{\begin{tabular}{ccccccc}
    \hline
       $K$ & $\eta(0)$  & $\dot{\eta}(0)$  & $\theta_m$ & $\sigma$ & $\altmathcal{L}$ \\
      \hline
       2.88  & 0.36 & -0.18 & 0.54 & 0.11 & 1.15\\
       1.14  & 0.05 & -0.01 & 0.49 & 0.10 & 0.05\\
       0.95  & 0.02 & -0.03 & 0.49 & 0.10 & 0.01\\
\hline 
\end{tabular}}
\label{tab:table1}
\end{table}

\section{Order parameter determined from experiments}
\label{appC}

The order parameter is calculated from the experimental data by assuming that the heat release rate fluctuations measured at each coarse-grained location comprise a set of limit-cycle oscillators \citep{popovych2005phase,bick2018chaos}. To minimize the effect of noisy fluctuations in the experimental measurements, we coarse-grain the unfiltered chemiluminescence images over $6\times6$ pixels at each state. We extract the time series at each state of combustor operation and then normalize the states by the amplitude $\dot{q}^\prime$ during the limit cycle oscillations. When the swirler rotation rate is progressively varied, the normalized time series at each coarse-grained location depicts a transition from high amplitude limit cycle oscillations to low amplitude chaotic oscillations.

In each image, we assume that each $m^\textrm{th}$ pixel consists of $n_k$ number of limit cycle oscillators. We then assume that $n_m=n$ for all the pixels to make our calculations easier. So, the phase for $j^\textrm{th}$ oscillator at $m^\textrm{th}$ pixel in the image be $\varphi_{mj}$, where $j$ varies as $1,..,n$. Now we can express, the complex order parameter for $m^\textrm{th}$ pixel as follows: $r_m e^{i \theta_m} = 1/n \sum_{m = 1}^{n} e^{i \varphi_{mj}}$ \citep{strogatz2000kuramoto}. We then calculate the absolute value and argument of the Hilbert transform of $\dot{q}^\prime_m(t)$, which is used for determining $r_m(t)$ and $\psi_m(t)$. Consequently, the expression for the order parameter obtained from experiments can be defined as:       
\begin{equation}
  \bar{r}e^{i\langle\theta\rangle} = \left\langle\frac{1}{N_p} \sum_{m=1}^{N_p}r_m e^{i \psi_m}\right\rangle_t,
  \label{eq:appC1}
\end{equation}
where consecutive averaging operations were executed over $N_p$ pixels in each image and the total number of images across the time series. The determined value of $\bar{r}$ is then used in figure \ref{fig:order_param}a. 

\bibliography{ref}

\end{document}